\documentstyle{article}

\newcounter{eqnletter}[equation]
\catcode`\@=11
\@addtoreset{equation}{section}   
\catcode`\@=12

\newcommand{\ea}{\setcounter{eqnletter}{1}\addtocounter{equation}{0}}
\newcommand{\eb}{\setcounter{eqnletter}{2}\addtocounter{equation}{-1}}
\newcommand{\ec}{\setcounter{eqnletter}{3}\addtocounter{equation}{-1}}
\newcommand{\ed}{\setcounter{eqnletter}{4}\addtocounter{equation}{-1}}
\newcommand{\ee}{\setcounter{eqnletter}{5}\addtocounter{equation}{-1}}

\begin{document}

KFT - UL - 4/94

\begin{center}

{\LARGE\bf Generalized Gaudin Models and \\[5mm]
Riccatians}

\vskip 1cm

{\large {\bf A.G. Ushveridze}}\footnote{This work was
partially supported by KBN grant no. 2P30221706p01}

\vskip 0.5 cm

Department of Theoretical Physics, University of Lodz,\\
Pomorska 149/153, 90-236 Lodz, Poland\footnote{E-mail
address: alexush@mvii.uni.lodz.pl and alexush@krysia.uni.lodz.pl} \\
(November 2, 1994)

\end{center}
\vspace{1 cm}
\begin{abstract}
The systems of differential
equations whose solutions exactly coincide with Bethe
ansatz solutions for generalized Gaudin models are constructed.
These equations are called the {\it generalized spectral
Riccati equations}, because the simplest equation of this
class has a standard Riccatian form. The general form of
these equations is
$R_{n_i}[z_1(\lambda),\ldots, z_r(\lambda)] = c_{n_i}(\lambda), \ i=1,\ldots,
r$, where $R_{n_i}$ denote some homogeneous polynomials of
degrees $n_i$ constructed from functional variables $z_i(\lambda)$
and their derivatives. It is assumed that $\deg \partial^k
z_i(\lambda) = k+1$. The problem is to find
all functions $z_i(\lambda)$ and $c_{n_i}(\lambda)$
satisfying the above equations under $2r$ additional constraints
$P \ z_i(\lambda)=F_i(\lambda)$ and $(1-P) \
c_{n_i}(\lambda)=0$, where $P$ is a projector from the space of
all rational functions onto the space of rational functions
having their singularities at {\it a priori} given points.
It turns out that this problem has solutions only for
very special polynomials $R_{n_i}$. Simplest
polynomials of such sort are called {\it Riccatians}.
One of most important results of the paper is the
observation that there exist one-to-one correspondence
between the systems of Riccatians and simple Lie algebras.
In particuler, the degrees of Riccatians associated with a
given simple Lie algebra ${\cal L}_r$ of rank $r$ coincide
with the orders of corresponding Casimir invariants.
In the paper we present an explicit form of Riccatians
associated with algebras $A_1, A_2, B_2, G_2, A_3, B_3, C_3$.
Another important result is that functions $c_{n_i}(\lambda)$
satisfying the system of generalized Riccati equations
constructed from Riccatians of the type ${\cal L}_r$
exactly coincide with eigenvalues of the Gaudin spectral problem
associated with algebra ${\cal L}_r$. This result suggests
that the generalized Gaudin models admit a total separation
of variables.

\vspace{1 cm} \end{abstract}

\section{Introduction}

The differential equations with large internal symmetries
always  have a great theoretical significance and, as a rule,
admit many interesting mathematical and physical applications.
One of such equations is the ordinary spectral Riccati equation\footnote{The
exact meaning of the adjective ``spectral'' will be clarified in subsection
1.1. Here we only note that the class of ordinary
spectral Riccati equations contains, for example, the
delinearized version of Lame equation.}
which is very well known to theoretical physicists interested
in properties of completely integrable quantum systems
and their solutions. The main feature of this equation is
that its solutions exactly coincide with
Bethe ansatz solutions of completely integrable quantum
Gaudin models associated with algebra $sl(2)$.

The aim of the present paper is to demonstrate that the ordinary
spectral Riccati equation admits a very natural
generalization to a multi-component case when instead of
a single first-order nonlinear differential equation
one considers the systems of nonlinear equations of higher orders.
It turns out that solutions of these systems
exactly coincide with Bethe ansatz solutions of generalized
Gaudin problems associated with various simple Lie algebras.

The paper is organized as follows. Here, in introducion,
we consider the simplest Riccati spectral problem
(subsection 1.1) and simplest Gaudin spectral problem
associated with algebra $sl(2)$ (subsection 1.2). In
subsection 1.3 we demonstrate the coincidence of
solutions of these problems
and present a standard explanation of this fact.

The main body of the paper is devoted to construction of generalized
Riccati spectral equations and their solutions. In section 2
we introduce all necessary notions and notations and
formulate some general theorems. In particular,
we introduce a new very important notion of {\it Riccatians}
which play the role of elementary building blocks by constructing
generalized Riccati equations.  In section 3 we discuss the
methods for calculating simplest Riccatians. The results
of these calculations are collected in next section 4.
In section 5 we demonstrate remarkable parallels between
systems of Riccatians and systems of
independent Casimir invariants for simple Lie algebras.
In the same section we give the final form of solution of
the generalized
Riccati spectral problem. In section 6 we consider the
generalized Gaudin models associated with arbitrary
simple Lie algebras and present their Bethe ansatz
solutions. In last section 7 we demonstrate the coincidence
of solutions of Riccati and Gaudin spectral problems and
discuss possible ways of its explanation.

\subsection{Simplest Riccati spectral problem}

Consider the followig formal relation
\begin{eqnarray}
z'(\lambda) + z^2(\lambda)=c(\lambda),
\label{1.1}
\end{eqnarray}
in which $z(\lambda)$ and $c(\lambda)$ are assumed to
be some analytic functions of a complex variable $\lambda$.

First of all note that the relation (\ref{1.1}) (if one
considers it as an equation) admits at
least two interpretations:

1. {\it Function $z(\lambda)$ is given, and function
$c(\lambda)$ is being sought}. This problem is trivial
and has a unique solution.

2. {\it Function $c(\lambda)$ is given, and function
$z(\lambda)$ is being sought}. This problem is nothing else
than the ordinary Riccati equation.
It has a one-parameter set of solutions. Except some very special
cases this problem cannot be solved in quadratures [Korn
and Korn 1971].

It turns out however that along with these two polar
interpretations, there exists an
interesting intermediate one which leads
to a very rich set of solutions and has a great theoretical
importance.  Roughly speaking, the idea of this
interpretation is to fix appropriately {\it some parts}
of {\it both} functions
$z(\lambda)$ and $c(\lambda)$ and state the problem of
finding the {\it remaining parts} of these functions. Before
giving a rigorous formulation of this problem, let us
introduce some necessary notions and notations.

Let ${\cal R}$ be the class of all rational functions $r(\lambda)$
of a single complex variable $\lambda$. This class can obviously be
viewed as an infinite-dimensional linear vector space with
a basis consisting of the so-called {\it elementary
rational functions}. We denote these functions by $r_a^n(\lambda)$
and define them as $(\lambda-a)^{-n}$, for $a\neq \infty$, and
as $\lambda^{n-1}$ for $a=\infty$. In both cases $n$ is a
natural number.

We call a rational function singular (regular) at the point
$\lambda=a$ if its expansion in elementary rational functions contains
(does not contain) a term proportional to $r_a^n(\lambda)$ with
some $n$. For example, according to our definition,
function $r(\lambda)=\lambda^{-1}$ is singular at the point
$\lambda=0$ and regular at all other points including infinity.
As to the function $r(\lambda) = 1$, it is singular at
infinity but regular at all finite points.

Let $A$ be the a finite set of nonequal fixed complex numbers (one of
which may be the infinity), and
$B$ be the set of all the remaining numbers. Denote by
${\cal R}_A$ and ${\cal R}_B$ the classes of those rational
functions from ${\cal R}$
whose singularities belong only to the sets $A$ and $B$, respectively.
Considering the classes ${\cal R}_A$ and ${\cal R}_B$ as
linear vector spaces we can write ${\cal R}_A\oplus{\cal R}_B={\cal R}$.
Denote also by ${\cal P}_A$ and ${\cal P}_B$ the projectors from ${\cal R}$
onto ${\cal R}_A$ and ${\cal R}_B$ for which we obviously
have ${\cal P}_A+{\cal P}_B=1$.

Now we are ready to give a strict formulation of the problem of our interest.

\medskip
{\bf Problem 1.1.} {\it Find all functions $z(\lambda) ,
c(\lambda) \in {\cal R}$
satisfying the relation (\ref{1.1}) under
two additional constraints:
\begin{eqnarray}
{\cal P}_A\ z(\lambda)=F(\lambda), \quad {\cal P}_B\ c(\lambda)=0,
\label{1.2}
\end{eqnarray}
in which $F(\lambda)\in {\cal R}_A$ is a given function.}

\medskip
The first constraint means that all the functions $z(\lambda)$
should have the same fixed projection $F(\lambda)$ onto the space ${\cal
R}_A$ and, in principle, may have arbitrary singularities outside the set $A$.
The second constraint means however that not any singularities of
functions $z(\lambda)$ lying outside the set $A$ are
admissible, but only those, at which the second function
$c(\lambda)$ is regular.

\medskip
{\bf Definition 1.1.} The expression $P[z(\lambda)]\equiv
z'(\lambda)+z^2(\lambda)$ standing in the left hand side of
formula (\ref{1.1}) we shall call the {\it Riccati
polynomial} of a functional variable $z(\lambda)$.
Functions $z(\lambda)$ and $c(\lambda)$ satisfying
the conditions of the problem 1.1 we
shall call the {\it eigenfunctions} and {\it eigenvalues} of
the Riccati polynomial $R[z(\lambda)]$.
The set  $\{z(\lambda), c(\lambda)\}$ of all such
eigenfunctions and eigenvalues we shall call
the {\it spectrum} of the Riccati polynomial, and the
problem 1.1 itself --- the {\it Riccati spectral problem}.

\medskip
In order to find the general solution of this problem, it is
reasonable to consider first the following auxiliary subproblem.
Assume that the rational function $z(\lambda)$ has
an isolated singularity at the point $\xi \not\in A$ and
try to find conditions under which the polynomial
$R[z(\lambda)]$ is regular at the point $\lambda=\xi$.

First of all note that the only possible singularity of the
function $z(\lambda)$ at the point $\lambda=\xi$ may be
simple pole because, as it can be easily verified, the
higher poles cannot be cancelled in the expression
for $R[z(\lambda)]$ at all. This allows one to
represent function $z(\lambda)$ in the form
\begin{eqnarray}
z(\lambda) = y(\lambda)+ \frac{v}{\lambda - \xi},
\label{1.3}
\end{eqnarray}
where $v$ is some complex number and $y(\lambda)$ is a
function regular at $\lambda=\xi$. Substituting (\ref{1.3})
into Riccati polynomial $R[z(\lambda)]$ we obtain
\begin{eqnarray}
R\left[y(\lambda)+ \frac{v}{\lambda-\xi}\right] =
\frac{R_0[v,y(\xi)]}{(\lambda-\xi)^2}+
\frac{R_1[v,y(\xi)]}{\lambda-\xi}+\mbox{regular trerms}.
\label{1.4}
\end{eqnarray}
The coefficients $R_0[v,y(\xi)]=v^2-v$ and
$R_1[v,y(\xi)]=2v y(\xi)$ we call the {\it residues} of
the polynomial $R[z(\lambda)]$. Now note that there exists
a special value of $v$, namely $v=1$, for which
both the residues become proportional to the function $y(\xi)$:
$R_0[1,y(\xi)]=0$ and $R_0[1,y(\xi)]=2 y(\xi)$. This means
that for $v=1$ the condition of regularity of
the polynomial $R[z(\lambda)]$
at the point $\lambda=\xi$ can be written as
\begin{eqnarray}
y(\xi)=0.
\label{1.5}
\end{eqnarray}
Formula (\ref{1.5}) enables one to present the general
solution of the problem 1.1.
Indeed, from the condition of the absence of higher order
poles in $z(\lambda)$ it follows that the most general form
of this function is
\begin{eqnarray}
z(\lambda) = F(\lambda)+ \sum_{i=1}^M\frac{v_i}{\lambda - \xi_i},
\label{1.6}
\end{eqnarray}
where $M$ is an arbitrarily fixed non-negative integer and
$\xi_i$ are some unknown parameters. Rewritting (\ref{1.6}) in
one of the following $M$ forms
\begin{eqnarray}
z(\lambda) = y_i(\lambda)+ \frac{v_i}{\lambda -
\xi_i}, \quad i=1,\ldots, M
\label{1.7}
\end{eqnarray}
and using (\ref{1.5}) we
can conclude that the conditions of regularity of the
function $R[z(\lambda)]$ at the points $\lambda=\xi_i$ are
\begin{eqnarray}
y_i(\xi_i)=0, \quad i=1, \ldots, M,
\label{1.8}
\end{eqnarray}
provided that $v_i=1, \ i=1,\ldots, M$.
Using the explicit form of fuctions $y_i(\lambda)$,
\begin{eqnarray}
y_i(\lambda)=
F(\lambda)+ \sum_{k=1, k \neq i}^M\frac{v_k}{\lambda - \xi_k},
\quad i=1,\ldots, M,
\label{1.9}
\end{eqnarray}
we can write down the final solution of the problem 1.1.

\medskip
{\bf Theorem 1.1.} {\it The most general solution of the
problem 1.1 has the following form
\begin{eqnarray}
z(\lambda) &=& F(\lambda)+ \sum_{i=1}^M\frac{1}{\lambda - \xi_i},
\label{1.10}\ea\label{1.10a}\\
c(\lambda) &=&
\left(F(\lambda) + \sum_{i=1}^M\frac{1}{\lambda-\xi_i}\right)'+
\left(F(\lambda) + \sum_{i=1}^M\frac{1}{\lambda-\xi_i}\right)^2,
\eb\label{1.10b}
\end{eqnarray}
where $M$ is an arbitrary non-negative integer and
the numbers $\xi_i, \ i=1,\dots,M$ satisfy the system of equations
\begin{eqnarray}
\sum_{k=1, k\neq i}^M\frac{1}{\xi_i - \xi_k}+F(\xi_i)=0, \quad
i=1,\ldots, M.
\label{1.11}
\end{eqnarray}
For any $M=0,1,\ldots$ the system (\ref{1.10}) has a finite
set of solutions. Therefore the spectrum of the Riccati
polynomial (\ref{1.1}) is infinite and discrete.}

The most interesting feature of these solutions is that their
exactly coincide with Bethe ansatz solutions of the so-called $sl(2)$
Gaudin spectral problem. But before demonstrating this fact, it is
reasonable to remind the reader what does the $sl(2)$
Gaudin problem mean.

\subsection{Simplest Gaudin spectral problem}

The Gaudin spectral problems can be formulated in terms of the so-called
Gaudin algebras which are some special infinite-dimensional
extensions of simple Lie algebras.
In this section we consider the simplest Gaudin algebra
${\cal G}[sl(2)]$ associated
with algebra $sl(2)$. The discussion of the general case
will be given in section 6.

The three generators of the Gaudin algebra ${\cal G}[sl(2)]$,
which we denote by $S^\pm(\lambda)$ and $S(\lambda)$, are
parametrized by a complex parameter $\lambda$ playing
the role of an additional continuous index. The
commutation relations for these generators have the form
\begin{eqnarray}
[S(\lambda),S^\pm(\mu)]=\pm
\frac{S^\pm(\lambda)-S^\pm(\mu)}{\mu-\lambda},
\
[S^+(\lambda),S^-(\mu)]=\frac{S(\lambda)-S(\mu)}{\mu-\lambda},
\label{a.12}
\end{eqnarray}
and can be viewed as generalizations of commutation
relations for algebra $sl(2)$ [Gaudin 1976, 1983].

The lowest weight representations of Gaudin algebra
${\cal G}[sl(2)]$ can be defined by the formulas
\begin{eqnarray}
S^-(\lambda)|0\rangle=0, \quad S(\lambda)|0\rangle=F(\lambda)|0\rangle.
\label{a.18}
\end{eqnarray}
As in the $sl(2)$ case, $|0\rangle$ is the lowest weight vector,
and $F(\lambda)$ is the corresponding lowest weight
which, however, is here a function rather than a constant. The
representation space is given by the formula
\begin{eqnarray}
W_{F(\lambda)}=\mbox{linear span of
vectors}\ \{S^+(\lambda_1)\cdots S^+(\lambda_n)|0\rangle\},
\  n=0,1,2,\ldots.
\label{a.19}
\end{eqnarray}
with arbitrary $\lambda_1, \ldots, \lambda_n$.
Generally, this space is infinite-dimensional.

Consider the operator
\begin{eqnarray}
C(\lambda)= S^2(\lambda)+S^+(\lambda)S^-(\lambda)+
S^-(\lambda)S^+(\lambda),
\label{a.14}
\end{eqnarray}
which belongs to the universal enveloping algebra of
algebra ${\cal G}[sl(2)]$ and has the form similar to
the form of the Casimir operator for algebra $sl(2)$.
For this reason we call (\ref{a.14}) the {\it
Casimir--Gaudin operator}.
It is not difficult to see, however, that this
operator {\it is not} a Casimir invariant for the algebra ${\cal
G}[sl(2)]$ because it explicitly depends on the continuous
index $\lambda$ which is not contracted. In fact, one can
check that $C(\lambda)$ does
not commute with the generators of Gaudin algebra. At the
same time, it has another remarkable property which
is the commutativity of its ``values'' for different values of $\lambda$:
\begin{eqnarray}
[C(\lambda),C(\mu)]=0.
\label{a.15}
\end{eqnarray}
This property suggests to interpret $C(\lambda)$ as a
generating function of commuting integrals of motion (hamiltonians) for
some quantum system. It can be shown that this system is
completely integrable in the sense that it admits enough
number of independent\footnote{The independence of quantum
integrals of motion is understood here as functional
independence of their classical analogs obtained after dequantization.}
commuting integrals of motion [Sklyanin 1991].

It turns out that this system is not only completely
integrable, but also exactly solvable\footnote{The notions
of complete integrability and exact solvability in quantum
mechanics do not necessarily coincide [Doebner and
Ushveridze 1994]}.
Considering the representation space of
Gaudin algebra ${\cal G}[sl(2)]$ as a space of states, we can
formulate the following analog of the Schr\"odinger problem.

\medskip
{\bf Problem 1.2.} {\it Find all solutions of the spectral equation
\begin{eqnarray}
C(\lambda)\phi=c(\lambda)\phi, \quad \phi \in W_{F(\lambda)},
\label{a.21}
\end{eqnarray}
provided that the lowest weight $F(\lambda)$ is given}.

\medskip
{\bf Definition 1.2.} The equation (\ref{a.21}) is called
the {\it $sl(2)$ Gaudin spectral equation} and
the models described by ``hamiltonians'' $C(\lambda)$ we
refer to as {\it $sl(2)$ Gaudin models}.

\medskip
{\bf Theorem 1.2.} {\it An explicit solution of the problem 1.2
does exist and has an elegant and purely algebraic form:
\begin{eqnarray}
\phi &=& S^+(\xi_1)\cdot\ldots\cdot S^+(\xi_M)|0\rangle,
\label{a.23}\ea\label{a.23a}\\
c(\lambda) &=& \left(F(\lambda) +
\sum_{i=1}^M\frac{1}{\lambda-\xi_i}\right)'+\left(F(\lambda)
+ \sum_{i=1}^M\frac{1}{\lambda-\xi_i}\right)^2.
\eb\label{a.23b}
\end{eqnarray}
Here $M$ is an arbitrary non-negative integer and
the numbers $\xi_1, \ldots, \xi_M$ satisfy the system of equations
\begin{eqnarray}
\sum_{k=1,k\neq i}^M\frac{1}{\xi_i - \xi_k}+F(\xi_i)=0, \quad
i=1,\ldots, M.
\label{a.24}
\end{eqnarray}
For any rational function $F(\lambda)$
and for any finite $M$ the set of equation (\ref{a.24}) is
finite. This means that the whole spectrum of the $sl(2)$
Gaudin model is infinite and discrete.}

\medskip
The substitution (\ref{a.23a}) solving the problem is
called the {\it Bethe ansatz}, and the equations (\ref{a.24}) are
known under generic name of {\it Bethe ansatz equations}.
The action of the operator $C(\lambda)$ on the trial Bethe
vector (\ref{a.23a}) leads to two groups of terms, one of
which are proportional to (\ref{a.23a}) and other are not.
The latter are often called the {\it unvanted terms},
and the equations (\ref{a.24}) are exactly
the conditions for their cancellation. The complete proof
of theorem 1.2 can be found in [Gaudin 1983].

\subsection{On equivalence of Riccati and Gaudin problems in
the simplest case}

Comparing formulas (\ref{1.10b}) and (\ref{1.11}) with (\ref{a.23b}) and
(\ref{a.24}), the reader can easily make sure
that solutions of Riccati and Gaudin
spectral problems exactly coincide! This suggest that there
should be some deep relationship between these problems.
Actually, it turns out that the coincidence of solutions of problems
1.1 and 1.2 is not accidental and the reason for it is
the {\it separability of variables} in the Gaudin spectral
equation [Sklyanin 1987].
Moreover, there exists a procedure (which is known
in the literature as the {\it inverse procedure of separation of
variables} which enables one to derive the Gaudin spectral
problem from the Riccati one [Ushveridze 1989, 1994]. The main idea of this
derivation is based on the fact that the Riccati equation
is {\it linearizable} by the substitution
$z(\lambda)=\psi'(\lambda)/\psi(\lambda)$ after which it
takes the form of the so-called {\it linear
multi-parameter spectral equations}, i.e. linear spectral equations containing
many spectral parameters. The number of spectral parameters
is finite if $F(\lambda)$ is a rational function. Moreover,
these parameters enter into equation {\it linearly}. It is
known that any linear multi-parameter spectral equation
with linear dependence on spectral parameters can be interpreted as an
equation appearing after separation of
variables in some multi-dimensional completely integrable
quantum system. The explicit construction of this system
shows that it is nothing else than the $sl(2)$ Gaudin model.

We do not intend to discuss in this paper the details of
this derivation because it can be found in references
[Ushveridze 1989, 1994].
We aimed only to stress the fact that equation (\ref{1.1})
contains in a hidden form the complete information of the $sl(2)$
Gaudin model and its solutions.

In next sections of this paper we construct the
generalizations of equation (\ref{1.1}) which will contain
all information of general Gaudin models (associated with arbitrary
simple Lie algebras) and their solutions.

\section{Generalized Riccati spectral problem}

\subsection{Preliminaries}

In this section we shall consider various
differential operators acting on functions
of a complex variable $\lambda$. The simplest operator of
such sort is the ordinary first-order differential operator
$\partial\equiv\partial/\partial \lambda$. We shall
consider it as a graded object, having, by definition, the
unit degree of homogeneity,
\begin{eqnarray}
\mbox{deg}\ \partial =1.
\label{g.1}
\end{eqnarray}
The functions on which the differential operators will act,
will be vector-valued complex analytic functions
${\bf z}(\lambda)=\{z_1(\lambda),\ldots,z_r(\lambda)\}$. These
functions also will be considered as graded objects, having,
by definition, a vector grading ${\bf
n}=\{n_1,\ldots,n_r\}$. We express this fact by writing
$\deg z_1(\lambda)=n_1,\ldots, \deg z_r(\lambda)=n_r$, or, in
vector notations,
\begin{eqnarray}
\mbox{deg}\ {\bf z}(\lambda)\equiv{\bf n}.
\label{g.2}
\end{eqnarray}
Below we shall always assume that the components of vector
${\bf n}$ are natural numbers ordered as $n_1\le\ldots\le n_r$.
We can also establish some order relations between
different vectors by writing ${\bf n}_1\le {\bf n}_2$ if
each component of ${\bf n}_1$ is equal or less than the corresponding
component of ${\bf n}_2$. In this sense, there exist a
minamal vector consisting of unit components only.
We denote it by ${\bf u}=\{1,\ldots, 1\}$. This vector will
not be used only as a grading vector but also as a tool for
simplification of various notations. So, for example, the sum of all
components of vector ${\bf n}$ we shall often write
in compact form as ${\bf u}{\bf \cdot}{\bf n}$.

Let  $\Phi({\bf n})$ denote a space of graded
vector-valued complex analytic functions ${\bf z}(\lambda)$ satisfying
the condition (\ref{g.2}).
Taking into account formula (\ref{g.1}),
we can formulate several obvious statements:

\medskip
{\bf 1.} Let ${\bf z}(\lambda)\in \Phi({\bf n})$.
Then $\partial {\bf z}(\lambda)\in \Phi({\bf n}+{\bf u})$.

{\bf 2.} Let ${\bf z}_1(\lambda),{\bf z}_2(\lambda)
\in \Phi({\bf n})$ and $c_1,c_2\in C$. Then
$c_1 {\bf z}_1(\lambda)+c_2 {\bf z}_2(\lambda)\in \Phi({\bf n})$.

{\bf 3.} Let ${\bf z}_1(\lambda)\in
\Phi({\bf n}_1)$ and
${\bf z}_2(\lambda)\in \Phi({\bf n}_2)$. Then
${\bf z}_1(\lambda)\otimes {\bf z}_2(\lambda)\in
\Phi({\bf n}_1\otimes {\bf n}_2)$.

{\bf 4.} Let ${\bf z}_1(\lambda)\in
\Phi({\bf n}_1)$ and
${\bf z}_2(\lambda)\in \Phi({\bf n}_2)$. Then
${\bf z}_1(\lambda)\oplus {\bf z}_2(\lambda)\in
\Phi({\bf n}_1\oplus {\bf n}_2)$.

{\bf 5.} Let ${\bf z}(\lambda)\in \Phi({\bf n})$, where
${\bf n}={\bf n}_1\oplus {\bf n}_2)$. Then
${\bf z}(\lambda)={\bf z}_1(\lambda)\oplus {\bf z}_2(\lambda)$,
where ${\bf z}_1(\lambda)\in \Phi({\bf n}_1)$ and
${\bf z}_2(\lambda)\in \Phi({\bf n}_2)$.

\medskip
Using these simple rules and starting with the elements
${\bf z}_1(\lambda)$ of the space $\Phi({\bf n}_1)$, we can
construct the elements ${\bf z}_2(\lambda)$ of another
space $\Phi({\bf n}_2)$.
In this case the components of vectors ${\bf z}_2(\lambda)$ will have the
form of {\it polynomials} in components of vectors ${\bf z}_1(\lambda)$
and their finite derivatives.

\medskip
{\bf Definition 2.1.} The non-linear differential operators
$P$ constructed according the rules 1 -- 5 and
realizing the mapping $\Phi({\bf n}_1)\rightarrow \Phi({\bf n}_2)$
with $\dim {\bf n}_1 = \dim {\bf n}_2=r$,
we shall call the {\it r-operators} of the type
$|{\bf n}_2\rangle\langle{\bf n}_1|$.

\medskip
Let $P_1$ and $P_2$ be two r-operators of the types
$|{\bf n}_1\rangle\langle{\bf m}_1|$ and $|{\bf n}_2\rangle\langle{\bf m}_2|$,
respectively. We call $P_1$ {\it compatible}
with $P_2$ if ${\bf n}_1={\bf m}_2$. In this case, it is possible to
construct a {\it composite} r-operator $P_2\circ P_1$ of the type
$|{\bf n}_2\rangle\langle{\bf m}_1|$.

Let us now introduce an important notion of r-determinants.
Consider a transformation of elements
${\bf z}_1(\lambda)\in \Phi({\bf n}_1)$ into elements
${\bf z}_2(\lambda)\in \Phi({\bf n}_2)$ realized by a certain r-operator
$P$:
\begin{eqnarray}
{\bf z}_2(\lambda)= P[{\bf z}_1(\lambda)].
\label{g.3}
\end{eqnarray}
Let $\hat{\bf z}_1(\lambda)=\{\partial^k {\bf z}_1(\lambda)\}_{k=1}^\infty$ and
$\hat{\bf z}_2(\lambda)=\{\partial^k {\bf z}_2(\lambda)\}_{k=1}^\infty$
denote some infinite-dimensional
vector functions of $\lambda$. Acting on both hand sides of (\ref{g.3})
by the operators $\partial^k$ with $k=0,1,2,\ldots, \infty$, one can construct
a new infinite-dimensional and non-differential operator $\hat
P$ realizing the transformation of vectors $\hat{\bf
z}_1(\lambda)$ into vectors $\hat{\bf z}_2(\lambda)$:
\begin{eqnarray}
\hat{\bf z}_2(\lambda)= \hat P[\hat {\bf z}_1(\lambda)].
\label{g.4}
\end{eqnarray}

\medskip
{\bf Definition 2.2.} The Jacobian of the transformation (\ref{g.4})
will be called the {\it r-determinant}
of the r-operator $P$ and denoted by $\mbox{r-det}\ P$.

\medskip
{\bf Lemma 2.1.} {\it The r-determinants
have all properties of ordinary determinants. In particular,
\begin{eqnarray}
\mbox{\rm r-det}\ (P_2\circ P_1)= (\mbox{\rm r-det}\ P_2) \cdot
(\mbox{\rm r-det}\ P_1),
\label{g.5}
\end{eqnarray}
for any r-operators $P_1$ and $P_2$}.

\medskip
{\bf Definition 2.3.} We call a r-operator $P$
{\it degenerate} if $\mbox{r-det}\ P=0$,
and {\it non-degenerate} if $\mbox{r-det}\ P\neq 0$.

\medskip
{\bf Lemma 2.2.} {\it Let $P$ be a r-operator of
the type $|{\bf n}\rangle\langle{\bf m}|$. If $P$ is
non-degenerate, then ${\bf n}\ge {\bf m}$}.

\medskip
{\bf Definition 2.4.} The non-degenerate r-operators of the
type $|{\bf n}\rangle\langle{\bf n}|$ we shall call {\it
pseudo-diagonal r-operators}.

\subsection{Riccati operators and polynomials}

Riccati operators are important particular cases of general
r-operators. Below we give their definition and discuss
some important properties.

\medskip
{\bf Definition 2.5.} The non-degenerate r-operator of the type
$|{\bf n}\rangle\langle{\bf u}|$ we shall call the {\it Riccati operator}
of dimension $r=\mbox{dim}\ {\bf u}$ and {\it age} $|{\bf n}\rangle$.
Let $R_1$ and $R_2$ be two Riccati operators of ages $|{\bf
n}_1\rangle$ and $|{\bf n}_2\rangle$. We call the operator
$R_1$ {\it older} ({\it not younger}) than $R_2$ or {\it youger}
({\it not older}) than $R_2$ if ${\bf n}_1>{\bf n}_2$
(${\bf n}_1\ge{\bf n}_2$) or ${\bf n}_1<{\bf n}_2$ (${\bf n}_1\le{\bf n}_2$),
respectively.

\medskip
Any Riccati operator $R^0$ of age $|{\bf n}^0\rangle$ is compatible with any
r-operator $P$ of age $|{\bf n}\rangle\langle{\bf n}^0|$.
If $P$ is non-degenerate, then the
composite operator $R=P\circ R$ is again a Riccati operator of
age $|{\bf m}\rangle$. According to lemma 2.2, the operator
$R$ is always not younger than $R^0$. Note also that pseudo-diagonal
transformations conserve the age of Riccati operators. The Riccati
operators connected by some pseudo-diagonal
transformation we shall call age-equivalent.

\medskip
{\bf Definition 2.6.} Let $R$ be a Riccati operator of dimension $r$ and age
$|{\bf n}\rangle$ acting in the space $\Phi({\bf u})$. If ${\bf z}(\lambda)
\in \Phi({\bf u})$, then $R[{\bf z}(\lambda)]\in \Phi({\bf n})$ is a
$r$-component vector function. Its components, which are the homogeneous
polynomials in components of function ${\bf
z}(\lambda)$ and their finite derivatives, we shall
denote by $R_{n_i}[{\bf z}(\lambda)], \ i=1,\ldots, r$ and call
the {\it Riccati polynomials} of degrees $n_i,\ i=1,\ldots, r$.

\medskip
As an example, below we present
the most general form of Riccati polynomials of
degrees $n=0,1,2,3$ and $4$.
\begin{eqnarray}
R_0[{\bf z}(\lambda)]&=&A_0,
\label{2.2}\ea\label{2.2a}\\
R_1[{\bf z}(\lambda)]&=&A_1\cdot {\bf z}(\lambda),
\eb\label{2.2b}\\
R_2[{\bf z}(\lambda)]&=&A_2\cdot \partial {\bf z}(\lambda)+
B_2\cdot {\bf z}(\lambda) \otimes {\bf z}(\lambda),
\ec\label{2.2c}\\
R_3[{\bf z}(\lambda)]&=&A_3\cdot \partial^2 {\bf z}(\lambda)+
B_3\cdot \partial {\bf z}(\lambda) \otimes {\bf z}(\lambda)+\nonumber\\
&+& C_3 \cdot {\bf z}(\lambda) \otimes {\bf z}(\lambda) \otimes {\bf
z}(\lambda)
,
\ed\label{2.2d}\\
R_4[{\bf z}(\lambda)]&=&A_4\cdot \partial^3 {\bf z}(\lambda)+
B_4\cdot \partial^2 {\bf z}(\lambda) \otimes {\bf z}(\lambda)+\nonumber\\
&+& C_4\cdot \partial {\bf z}(\lambda) \otimes \partial {\bf z}(\lambda)
+D_4 \cdot \partial {\bf z}(\lambda)
\otimes {\bf z}(\lambda) \otimes {\bf z}(\lambda) +\nonumber\\
&+& E_4 \cdot {\bf z}(\lambda) \otimes {\bf z}(\lambda)
\otimes {\bf z}(\lambda) \otimes {\bf z}(\lambda).
\ee\label{2.2e}
\end{eqnarray}
Here ``$\otimes$'' means the ordinary tensor product and
``$\cdot$'' denotes contraction of tensors of equal rank.
$A_n, B_n, C_n, D_n$ and $E_n$ are constant tensors of rank $n$.

\subsection{Riccati spectral problem}

Now we can start the discussion of spectral
properties of Riccati operators.
Let ${\bf z}(\lambda)\in \Phi({\bf u})$ and
${\bf c}(\lambda)\in \Phi({\bf n})$
be two analytic functions satisfying the
following relation:
\begin{eqnarray}
R[{\bf z}(\lambda)] = {\bf c}(\lambda),
\label{2.3}
\end{eqnarray}
where $R$ is a certain Riccati operator
of age $|{\bf n}\rangle$. This relation can
be interpreted as a generalization of the relation (\ref{1.1}).

As in the case of (\ref{1.1}) the relation
(\ref{2.3}) (if one considers it as an equation) admits two
standard interpretations:

1. {\it Functions ${\bf z}(\lambda)$ are given, and functions
${\bf c}(\lambda)$ are being sought}.
This problem is trivial and has a unique solution.

2. {\it Functions ${\bf c}(\lambda)$
are given, and functions
${\bf z}(\lambda)$ are being sought}. This problem can be
considered as a generalization of the ordinary Riccati equation.
It has a ${\bf u}{\bf \cdot}({\bf n}-{\bf u})$-parameter set of solutions.
Generally, this problem cannot be solved in quadratures.

In this section we discuss the third (intermediate) case
which is realized when {\it some parts}
of {\it both sets} of functions
${\bf z}(\lambda)$ and ${\bf c}(\lambda)$ are
simultaneously given and
the {\it remaining parts} of these functions are being sought.
It is reasonable to change a little bit the meaning of
notations introduced in section 1. Hereafter we denote by ${\cal R}$
the class of all $r$-component vector-valued rational functions
of a single complex variable $\lambda$. We denote also by
${\cal R}_{A}$ and ${\cal R}_{B}$ the classes of those rational
functions from ${\cal R}$ whose singularities belong only
to the sets $A$ and $B$ defined in section 1.
As before,
${\cal P}_{A}$ and ${\cal P}_{B}$ will be the projectors from ${\cal R}$
onto ${\cal R}_{A}$ and ${\cal R}_{B}$. Then the
problem of our interest can be formulated as follows.

\medskip
{\bf Problem 2.1.} {\it Find all functions ${\bf z}(\lambda) ,
{\bf c}(\lambda) \in {\cal R}$
satisfying the relation (\ref{2.3}) under
two additional constraints:
\begin{eqnarray}
{\cal P}_{A}\ {\bf z}(\lambda)={\bf F}(\lambda), \quad
{\cal P}_{B}\ {\bf c}(\lambda)=0,
\label{2.4}
\end{eqnarray}
in which ${\bf F}(\lambda)\in {\cal R}_A$ is a given vector
function.}

\medskip
{\bf Definition 2.7.} Functions ${\bf z}(\lambda)$
and ${\bf c}(\lambda)$ satisfying
the conditions of the problem we
shall call the {\it eigenfunctions} and {\it eigenvalues} of
the Riccati operator $R$.
The set  $\{{\bf z}(\lambda), {\bf c}(\lambda)\}$ of all such
eigenfunctions and eigenvalues we shall call
the {\it spectrum} of the Riccati operator $R$, and the
problem 2.1 itself we call the {\it (generalized) Riccati spectral problem}.

\medskip
Below we will have many opportunities to demonstrate that
not for any Riccati operator the solvability of the problem
2.1 can be guaranteed.

\medskip
{\bf Definition 2.8.} We call a Riccati operator
{\it admissible} if its spectrum is non-empty.

\medskip
The following problem immediately arises:

\medskip
{\bf Problem 2.2.} {\it Find all admissible Riccati operators}.

\subsection{Residues and regularizators}

In order to solve the problem 2.2 in full generality, it is
reasonable to choose the same strategy as in section 1 and
consider first the following auxiliary subproblem.

Assume that the rational function $ {\bf z}(\lambda)$ has
an isolated singularity at the point $\xi \not\in A$ and
try to find conditions under which the function
${\bf c}(\lambda) = R[{\bf z}(\lambda)]$ is regular at the point $\lambda=\xi$.

It is easy to check that the only possible singularity of the
function $ {\bf z}(\lambda)$ at the point $\lambda=\xi$ may be
a simple pole because the higher poles cannot be
cancelled in the expression for ${\bf c}(\lambda)$ at all. This allows one to
represent function ${\bf z}(\lambda)$ in the form
\begin{eqnarray}
{\bf z}(\lambda) =  {\bf y}(\lambda)+ \frac{ {\bf v }}{\lambda - \xi},
\label{2.5}
\end{eqnarray}
where $ {\bf v }$ is some  complex vector and ${\bf y}(\lambda)$
is a vector-valued function regular at $\lambda=\xi$.
Acting on (\ref{2.5})
by a certain Riccati operator $R$,
we obtain a vector-valued function whose components can be
represented in the form:
\begin{eqnarray}
R_{n_i}\left[ {\bf y}(\lambda)+
\frac{ {\bf v }}{\lambda-\xi}\right]=
\sum_{k=0}^{n_i-1} \frac{R_{ik}[{\bf v },
{\bf y}(\xi)]}{(\lambda-\xi)^{n_i-k}}+
\mbox{regular terms}, \quad i=1,\ldots,r.
\label{2.6}
\end{eqnarray}
Here  $R_{ik}[{\bf v },  {\bf y}(\xi)]$ are some vector
functions of ${\bf v }$
and $\partial^l {\bf y}(\xi), \ l\ge 0$.
It is not difficult to show that these functions are
some Riccati polynomials of degrees $k=0,1,\ldots,n_i-1$. Indeed,
for the equality $\mbox{deg} \ \partial^l  {\bf z}(\lambda)=l+1$
to be compatible with
(\ref{2.5}) it is necessary to take $\mbox{deg}
\  {\bf y}(\lambda)={\bf u}$ and $\mbox{deg}\ (\lambda-\xi)=-1$, provided that
$\mbox{deg}\ {\bf v }=0$. The fact that both hand sides of
(\ref{2.6}) should have the same degrees $n_i$ implies that
\begin{eqnarray}
\mbox{deg}\ R_{ik}[{\bf v }, {\bf y}(\xi)]=k.
\label{2.7}
\end{eqnarray}

\medskip
{\bf Definition 2.9.} The polynomials $R_{ik}[{\bf v },
{\bf y}(\xi)]$ of degrees $k=0,\ldots, n_{i_1}$ playing the role of
the coefficients for the singular terms in the expansion
(\ref{2.6}), we shall call the {\it residues} of the the
Riccati operator $R$.

\medskip
{}From formula (\ref{2.6}) it follows that the condition of
regularity of function ${\bf c}(\lambda)]$
at the point $\lambda=\xi$ is the
condition of a simultaneous vanishing of all the residues
of $R$:
\begin{eqnarray}
R_{ik}[{\bf v},{\bf y}(\xi)]=0, \quad
k=0,1,\ldots, n_i-1, \quad i=1,\ldots,r.
\label{2.8}
\end{eqnarray}
The total number of these conditions, equal to the number
of all residues, is obviously ${\bf u}{\bf \cdot}{\bf n}$,
while the number of unknowns in (\ref{2.8}), consisting of the components
of vector ${\bf v}$ and parameter $\xi$, is $\mbox{dim}\
{\bf u}+1$. In all the cases when ${\bf i\cdot
n}>\mbox{dim}\ {\bf u}+1$
the system (\ref{2.8}) is over-determined and thus has no solutions at all.

Assume, however, that there exist some special values of ${\bf v}$
for which all ${\bf u}{\bf \cdot}{\bf n}$  residues of
$R$ become proportional to a
{\it single} Riccati polynomial  of degree $l$:
\begin{eqnarray}
&R_{ik}[{\bf v},{\bf y}(\xi)]=Q_l[{\bf y}(\xi)]
R'_{i,k-l}[{\bf v},{\bf y}(\xi)],& \nonumber\\
&k=0,1,\ldots,n_i-1, \quad i=1,\ldots, r.&
\label{2.9}
\end{eqnarray}
In this case the system of conditions (\ref{2.8}) can be
reduced to a single equation
\begin{eqnarray}
Q_l[{\bf y}(\xi)]=0
\label{2.10}
\end{eqnarray}
for a single unknown $\xi$. Obviously, in general case
such systems are solvable.

\medskip
{\bf Definition 2.10.} The vector ${\bf v}$ for
which the residues $R_{ik}[{\bf v},{\bf y}(\xi)]$
of a Riccati operator $R$
become divisible (without remainder)
by a certain Riccati polynomial  $Q_l[{\bf y}(\xi)]$ of degree $l$,
we call the {\it $l$ - regularizing vector}, and the polynomial
$Q_l[{\bf y}(\xi)]$  itself we call the {\it $l$ - regularizing polynomial}.
The set of pairs $\{{\bf v}, Q_l[{\bf y}(\xi)]\}$
consisting of all $l$ - regularizing vectors and
corresponding $l$ - regularizing polynomials,
we call the {\it $l$ - regularizator} of the
Riccati operator $R$ and denote it by
$\mbox{reg}\ R$. The Riccati operators
with non-empty $l$ - regularizators
we call {\it $l$ - regularizable.}

\medskip
{\bf Conjecture 2.1.} {\it For any Riccati operator of dimension
$r$, the $l$ - regularizator consists of exactly
$r$ elements}.

\medskip
At the present time we have no satisfactory proof of this conjecture,
however, the experience accumulated by studying various
concrete Riccati operators indicates that it should be true.
At least it is true for all Riccati operators which we intend
to discuss in the present paper.

\subsection{General theorem}

Now we are ready to formulate the following theorem.

\medskip
{\bf Theorem 2.1.} {\it Let $R$
be a given Riccati operator of dimension $r$
with a non-empty reqularizator
$\{{\bf v}^a ,Q_l^a[{\bf y}(\xi)]\}, \ a=1,\ldots, r$.
Then solution of the
corresponding Riccati spectral problem 2.1 has the following form:
\begin{eqnarray}
{\bf z}(\lambda) &=& {\bf F}(\lambda)+
\sum_{a}\sum_{i=1}^{M_a}\frac{{\bf v}^a}
{\lambda - \xi_{i}^a},
\label{2.11}\ea\label{2.11a}\\
{\bf c}(\lambda) &=& R
\left[{\bf F}(\lambda)+ \sum_{a}\sum_{i=1}^{M_a}\frac{{\bf v}^a}
{\lambda - \xi_{i}^a}\right],
\eb\label{2.11b}
\end{eqnarray}
where $M_a, \ a=1,\ldots, r$ are arbitrary non-negative
integers, and the parameters $\xi_{i}^a, \ i=1,\ldots, M_a,
\ a=1,\ldots, r$ satisfy the system of equations
\begin{eqnarray}
Q_l^a\left[{\bf F}(\xi_{i}^a)+
\sum_{b}\sum_{k=1}^{M^b}\frac{{\bf v}^b}
{\xi_{i}^a - \xi_{k}^b}\right]=0, \quad i=1,\ldots, M_a,
\ a=1,\ldots, r.
\label{2.12}
\end{eqnarray}
For any set of non-negative integers $M_a, \ a=1,\ldots, r$ the
equations (\ref{2.12}) have a non-empty (finite)
set of solutions, and thus, the
Riccati operator under consideration is admissible and has an
infinite and discrete spectrum.}

\medskip
{\bf Proof.} The proof of this theorem is essentially the
same as the proof of the theorem 1.1 of section 1.
Indeed, using the same reasonongs as in section 1, we can
conclude that the most general form of function ${\bf z}(\lambda)$ satisfying
the constraints (\ref{2.4}) is (\ref{2.11a}),
where $M_a$ are some arbitrarily fixed non-negative
integers, ${\bf v}^a$ are arbitrary vectors and
$\xi_{i}^a$ are some complex parameters. Rewritting (\ref{2.13}) in
one of the following $M_1+\ldots+M_r$ forms
\begin{eqnarray}
{\bf z}(\lambda) = {\bf y}_{i}^a(\lambda)+ \frac{{\bf v}_{i}^a}{\lambda -
\xi_{i}^a}, \quad i=1,\ldots, M_a, \quad a=1,\ldots, r,
\label{2.13}
\end{eqnarray}
and using (\ref{2.10}), we
can conclude that the conditions of regularity of the
function ${\bf c}(\lambda)]$ at the points $\lambda=\xi_i^a$ are
\begin{eqnarray}
Q_l^a[{\bf y}_{i}^a(\xi_{i}^a)]=0, \quad i=1, \ldots, M_a,\quad
a=1,\ldots, r,
\label{2.14}
\end{eqnarray}
provided that ${\bf v}^{a}$ are regularizing vectors.
Using the explicit form of functions ${\bf y}_i^a(\lambda)$,
\begin{eqnarray}
{\bf y}_{i}^a(\lambda)=
{\bf F}(\lambda)+
\sum_{b}\sum_{k=1}^{M^b}\frac{{\bf v}^b}
{\lambda - \xi_{k}^b},
\quad i=1,\ldots, M_a, \quad a=1,\ldots, r,
\label{2.15}
\end{eqnarray}
we can write down the final solution (\ref{2.11}), (\ref{2.12})
of the problem 2.1. This completes the proof of the theorem.

\medskip
This theorem enables one to reduce the problem 2.2 of
finding all admissible Riccati operators to a more
simple and concrete one:

\medskip
{\bf  Problem 2.3} {\it Find all $l$ - regularizable Riccati
operators and construct their $l$ - regularizators}.

\medskip
Before trying to solve this problem it is useful to
introduce a very important notion of {\it Riccatians}.
This will be done in next subsection.

\subsection{Riccatians}

We start with two simple lemmas which enable one to
estimate the measure of ambigouity in fixing Riccati operators
by their $l$ - regularizators.

\medskip
{\bf Lemma 2.1.} {\it Let $R^0$ be some $l$ - regularizable
Riccati operator of age $|{\bf n}^0\rangle$,
compatible with a certain r-polynomial $P$ of type
$|{\bf n}\rangle\langle{\bf n}^0|$.
Then the composite Riccati operator $R=P\circ R^0$ of age $|{\bf n}\rangle$
is also $l$ - regularizable and
$\mbox{reg}\ R=\mbox{reg}\ R^0$. }

\medskip
{\bf Proof.} Let ${\bf v}$ and  $Q_l[{\bf y}(\lambda)]$ be
some $l$ - regularizing vector and
corresponding $l$ - regularizing polynomial.
Using formulas of the previous section,
we can write
\begin{eqnarray}
R^0[{\bf z}(\lambda)]=R^0\left[{\bf y}(\lambda)
+\frac{{\bf v}}{\lambda-\xi}\right]=Q_l[{\bf y}(\xi)]\cdot{\bf
s}(\lambda)+{\bf r}(\lambda),
\label{3.2}
\end{eqnarray}
where ${\bf s}(\lambda)$ and $r(\lambda)$ denote some
functions of $\lambda$, which, respectively, are singular and regular
at the point $\lambda=\xi$.
Taking the derivative of the both hand sides of formula
(\ref{3.2}), we obtain
\begin{eqnarray}
\partial R^0[{\bf z}(\lambda)]=\partial R^0\left[{\bf y}(\lambda)
+\frac{{\bf v}}{\lambda-\xi}\right]=Q_l[{\bf y}(\xi)]\cdot
\partial {\bf s}(\lambda)+\partial {\bf r}(\lambda).
\label{3.3}
\end{eqnarray}
We see that the singular part of this expansion is also proportional to
$Q_l[{\bf y}(\xi)]$.

Let now $R_1^0$ and $R_2^0$ be two Riccati operators
for which ${\bf v}$ and $Q_l[{\bf y}(\xi)]$
are known to be $l$ - regularizing vector and $l$ - regularizing
polynomial. Then we can write
\begin{eqnarray}
R_1^0[{\bf z}(\lambda)]=R_1^0\left[{\bf y}(\lambda)
+\frac{{\bf v}}{\lambda-\xi}\right]=Q_l[{\bf y}(\xi)]\cdot{\bf
s}_1(\lambda)+{\bf r}_1(\lambda),\label{3.4}\ea\label{3.4a}\\
R_2^0[{\bf z}(\lambda)]=R_2^0\left[{\bf y}(\lambda)
+\frac{{\bf v}}{\lambda-\xi}\right]=Q_l[{\bf y}(\xi)]\cdot {\bf
s}_2(\lambda)+{\bf r}_2(\lambda),
\eb\label{3.4b}
\end{eqnarray}
and, consequently,
\begin{eqnarray}
&&R_1^0[{\bf z}(\lambda)]\otimes R_2^0[{\bf z}(\lambda)]
=R_1^0\left[{\bf }(\lambda)+\frac{{\bf v}}{\lambda-\xi}\right]
\otimes R_2^0\left[{\bf }(\lambda)+\frac{{\bf v}}{\lambda-\xi}\right]
=\nonumber\\
&&Q_l^2[{\bf y}(\xi)]\cdot{\bf s}_1(\lambda)\otimes {\bf
s}_2(\lambda)+\nonumber
\\
&&Q_l[{\bf y}(\xi)]\cdot({\bf s}_1(\lambda)\otimes {\bf r}_2(\lambda)+
{\bf r}_1(\lambda)\otimes {\bf s}_2(\lambda))
+\nonumber\\
&&{\bf r}_1(\lambda)\otimes {\bf r}_2(\lambda).
\label{3.5}
\end{eqnarray}
We see that the singular part of this expression is, as before,
proportional to $Q_l[{\bf y}(\xi)]$.
Because the construction of the Riccati operator  $R$
implies the use of only two above operations of
differentiation and multiplication,
we can conclude that the singular part of the final expression for
$R[{\bf z}(\lambda)]$ will again be proportional to
$Q_l[{\bf y}(\lambda)]$. This means that the system of composite
Riccati polynomials should have at least the same $l$ - regularizator as the
initial system. According to conjecture 2.1,
these $l$ - regularizators should coincide.

\medskip
Lemma 3.1 demonstrates that $l$ - regularizator does not
fix a Riccati operator uniquely. There is always possible to start
with a given Riccati operator and construct another Riccati operator with the
same $l$ - regularizator but having a greater age.

\medskip
{\bf Definition 2.11.} The Riccati operators, having the minimal age
among all other Riccati operators with the same $l$ - regularizator,
we shall call {\it minimal} Riccati operators.

\medskip
It is not difficult to see that minimal Riccati operators
also cannot be uniquely determined by their $l$ - regularizators,
because of the existence of age conserving pseudo-diagonal transformations.

\medskip
{\bf Lemma 2.2.} {\it Let $R^1$ and
$R^2$ be two $l$ - regularizable Riccati operators.
Let a new Riccati operator $R$ be defined by the formula
\begin{eqnarray}
R[{\bf z}(\lambda)]=R[{\bf z}^1(\lambda)]\oplus
R^2[{\bf z}^2(\lambda)]
\label{3.8}
\end{eqnarray}
in which ${\bf z}(\lambda)={\bf z}^1(\lambda)\oplus{\bf z}^2(\lambda)$.
Then $R$ is $l$ - regularizable and
$\mbox{reg}\ \{R[{\bf z}(\lambda)]\}\approx
\mbox{reg}\ \{R^1[{\bf z}^1(\lambda)]\}\cup
\mbox{reg}\ \{R^2[{\bf z}^2(\lambda)]\}$}.

\medskip
{\bf Proof.} Let ${\bf v}^1$ and ${\bf v}^2$ be $l$ - regularizing
vectors and $Q_l^1[{\bf y}^1(\lambda)]$ and $Q_l^2[{\bf y}^2(\lambda)]$
be the corresponding $l$ - regularizing
polynomials for Riccati operators $R^1$ and $R^2$. Then we can write
\begin{eqnarray}
R^1[{\bf z}(\lambda)]=R^1\left[{\bf y}^1(\lambda)
+\frac{{\bf v}^1}{\lambda-\xi}\right]=Q_l^1[{\bf
y}^1(\xi)]\cdot {\bf
s}^1(\lambda)+{\bf r}^1(\lambda),\label{3.7}\ea\label{3.7a}\\
R^2[{\bf z}(\lambda)]=R^2\left[{\bf y}^2(\lambda)
+\frac{{\bf v}^2}{\lambda-\xi}\right]=Q_l^2[{\bf y}^2(\xi)]\cdot{\bf
s}^2(\lambda)+{\bf r}^2(\lambda).
\eb\label{3.7b}
\end{eqnarray}
{}From these formulas it immediately follows that
the Riccati operator $R$ has two $l$ - regularizing
vectors ${\bf v}^1 \oplus {\bf 0}$ and ${\bf 0}\oplus{\bf v}^2$ and two
corresponding $l$ - regularizing polynomials $Q_l^1[{\bf y}(\xi)]
=Q_l^1[{\bf y}^1(\xi)\oplus {\bf 0}]$  and
$Q_l^2[{\bf y}(\xi)]
=Q_l^2[{\bf 0}\oplus{\bf y}^2(\xi)]$. This completes the proof.

\medskip
{\bf Definition 2.12.} If a $l$ - regularizable Riccati operator $R$
can be constructed from two or more $l$ - regularizable
Riccati operators by formula (\ref{3.8}), we call this operator
{\it reducible}. Otherwise, we call it {\it irreducible}.

\medskip
Now we are ready to introduce a very important notion of Riccatians.

\medskip
{\bf Definition 2.13.} The minimal and irreducible $l$ - regularizable Riccati
operator we shall call {\it $l$-simple}.
The particular Riccati
polynomials associated with $l$-simple Riccati operators we call {\it
Riccatians of genus $l$} or, simply, {\it Riccatians}.

\medskip
Now we see that in order to solve problem 2.2 it is
sufficient to solve the follewin auxiliaty  problem:

\medskip
{\bf Problem 2.4.} {\it Find all systems of Riccatians.}

\medskip
In next section we start discussion of this problem
restricting ourselves by looking only for Riccatians of
genus 1.

\section{Construction of Riccatians}

This section is devoted to explicit construction of some simplest
systems of Riccatians. We start with most general expressions of Riccati
operators of {\it a priori} given age and, using general prescriptions given
in previous section, try to find those forms of these
operators for which they become simple. Here we
analyze from this point of view three general Riccati operators
of ages $|2\rangle$, $|2,2\rangle$ and $|2,3\rangle$.

\subsection{Riccati operators of age $|2\rangle$}

The set of Riccati operators of age $|2\rangle$ is
defined by the formula
\begin{eqnarray}
R_2[z(\lambda)]=A_2\partial z(\lambda) + B_2 z^2(\lambda).
\label{k.1}
\end{eqnarray}
in which $A_2$ and $B_2$ are arbitrary numerical parameters.
In order to find for which values of these parameters the
polynomial $R_2[z(\lambda)]$ becomes Riccatian, we should
analyze its residues and construct corresponding regularizator.

{}From general formulas of previous section it follows that
the operator (\ref{k.1}) has two residues
\begin{eqnarray}
R_{2,0}[y(\xi)]&=&-A_2v+B_2 v^2,
\label{k.4}\ea\label{k.4a}\\
R_{2,1}[y(\xi)]&=&2B_2vy(\xi),
\eb\label{k.4b}
\end{eqnarray}
and conditions for these residues to have a common divisor
of first degree
\begin{eqnarray}
Q[y(\xi)]=Q\cdot y(\xi)
\label{k.5}
\end{eqnarray}
are
\begin{eqnarray}
&-A_2v+B_2v^2=0,&
\label{k.6}\ea\label{k.6a}\\
&B_2v=e_2Q.&
\eb\label{k.6b}
\end{eqnarray}
Here $e_2$ is some arbitrary constant. We see that for any
choice of coefficients $A_2$ and $B_2$, there exists such
value of $e_2$ for which system (\ref{k.6}) has a single non-zero
solution for $v$. So that all operators (\ref{k.6}) are regularizable.
Because operators (\ref{k.1}) are also irreducible and
non-minimizible, we can assert that they are simple and polynomials
(\ref{k.1}) are Riccatians for any non-zero values of $A_2$
and $B_2$.

In order to find the canonical form of these Riccatians, we
should take
\begin{eqnarray}
v=1.
\label{k.7}
\end{eqnarray}
Substituting (\ref{k.7}) into (\ref{k.6}), we find that
$A_2=B_2$. Without loss of generality we can also take
$A_2=B_2=1$ which corresponds to a special choice of normalization.
This results in the following final expressions:
\begin{eqnarray}
R_2[z(\lambda)]=\partial z(\lambda) + z^2(\lambda),
\label{k.8}
\end{eqnarray}
for the Riccatian and
\begin{eqnarray}
v=1, \quad Q[y(\xi)]=y(\xi),
\label{k.9}
\end{eqnarray}
for its regularizator.

\subsection{Riccati operators of age $|2,2\rangle$}

Let us now consider Riccati operators of age
$|2,2\rangle$ the most general form of which is given
by formulas
\begin{eqnarray}
R_2[{\bf z}(\lambda)]=A^i_2\partial z_i(\lambda) +
B^{ik}_2z_i(\lambda)z_k(\lamb
da),
\label{l.1}\ea\label{l.1a}\\
\bar R_2[{\bf z}(\lambda)]=\bar A^i_2\partial z_i(\lambda) +
\bar B^{ik}_2z_i(\lambda)z_k(\lambda),
\eb\label{l.1b}
\end{eqnarray}
in which $A_2^{i}, B_2^{ik}$ and $\bar A_2^{i}, \bar B_2^{ik}$,
are some arbitrary parameters. The indices $i$ and $k$ in
(\ref{l.1}) take the values $1$ and $2$ and
the summation over repeated indices is assumed.

As before, in order to find the values of parameters for
which the polynomials (\ref{l.1a}) and (\ref{l.1b}) become
Riccatians, we should first look at the residues of
these polynomials
\begin{eqnarray}
R_{2,0}[{\bf y}(\xi)]&=&-A^i_2v _i+B^{ik}_2v _iv _k,
\label{l.4}\ea\label{l.4a}\\
R_{2,1}[{\bf y}(\xi)]&=&2B^{ik}_2v _iy_k(\xi),
\eb\label{l.4b}
\end{eqnarray}
and
\begin{eqnarray}
\bar R_{2,0}[{\bf y}(\xi)]&=&-\bar A^i_2v _i+\bar B^{ik}_2v _iv _k,
\label{l.5}\ea\label{l.5a}\\
\bar R_{2,1}[{\bf y}(\xi)]&=&2\bar B^{ik}_2v _iy_k(\xi).
\eb\label{l.5b}
\end{eqnarray}
The conditions for all these residues to have a common
divisor of degree one,
\begin{eqnarray}
Q[y(\xi)]=Q^ly_l(\xi),
\label{l.6}
\end{eqnarray}
are, respectively,
\begin{eqnarray}
&-A^i_2v _i+B^{ik}_2v _iv _k=0,&
\label{l.7}\ea\label{l.7a}\\
&B^{li}_2v _i=e_2Q^l&
\eb\label{l.7b}
\end{eqnarray}
and
\begin{eqnarray}
&-\bar A^i_2v _i+\bar B^{ik}_2v _iv _k=0,&
\label{l.8}\ea\label{l.8a}\\
&\bar B^{li}_2v _i=\bar e_2Q^l,&
\eb\label{l.8b}
\end{eqnarray}
where $e_2$ and $\bar e_2$ are
some non-zero numbers.

Comparing formulas (\ref{l.7b}) and (\ref{l.8b}), we can write
\begin{eqnarray}
\frac{1}{e_2}B^{li}_2v _i=\frac{1}{\bar e_2}\bar B^{li}_2v _i
\label{l.9}
\end{eqnarray}
or, in matrix notations,
\begin{eqnarray}
B_2\bar B_2^{-1}{\bf v} =\epsilon_2{\bf v},
\label{l.10}\ea\label{l.10a}\\
\bar B_2 B_2^{-1}{\bf v} =\bar \epsilon_2{\bf v},
\eb\label{l.10b}
\end{eqnarray}
where $\epsilon=e_2\bar e_2^{-1}$ and $\bar \epsilon=\bar e_2 e_2^{-1}$.
Relations (\ref{l.10}) mean that vectors ${\bf v}$ are the
eigenvectors of the matrices $B_2\bar B_2^{-1}$ and $\bar B_2 B_2^{-1}$,
while the numbers $\epsilon_2$ and $\bar\epsilon_2$ are their eigenvalues.
Because $B_2\bar B_2^{-1}$ and $\bar B_2 B_2^{-1}$ are $2\times 2$ matrices,
the equations (\ref{l.10}) should have at least two linearly
independent solutions which we
denote by ${\bf v}^n=\{v^n_i\}$.
The linear independence of these vectors enables one to
choose such a basis in two-dimensional vector space, in
which they become orthonormal repers
\begin{eqnarray}
v^n_i=\delta^n_i.
\label{l.11}
\end{eqnarray}
Denoting the corresponding eigenvalues by $\epsilon_2^n$ and
$\bar \epsilon_2^n$ and substituting (\ref{l.11}) into (\ref{l.10}),
we obtain
\begin{eqnarray}
B^{ln}_2&=&\epsilon^n\bar B^{ln}_2,
\label{l.12}\ea\label{l.12a}\\
\bar B^{ln}_2&=&\bar \epsilon^n B^{ln}_2.
\eb\label{l.12b}
\end{eqnarray}
Let us now remember that both matrices $B^{ln}_2$ and $\bar
B^{ln}_2$ are symmetric. Permuting the indices $l$ and $m$
in (\ref{l.12}) and subtracting obtained relations from
each other we obtain
\begin{eqnarray}
(\epsilon^n-\epsilon^l)\bar B^{ln}_2&=&0,
\label{l.13}\ea\label{l.13a}\\
(\bar \epsilon^n-\bar \epsilon^l) B^{ln}_2&=&0.
\eb\label{l.13b}
\end{eqnarray}
These relations can be satisfied in two cases: 1) if $\epsilon^1=\epsilon^2=
\varepsilon$
(or $\bar\epsilon^1=\bar\epsilon^2=\varepsilon$) and $B^{ln}_2$
(or $\bar B^{ln}_2$) is arbitrary, and 2) if $\epsilon^1\neq\epsilon^2$ (or
$\bar\epsilon^1\neq\bar\epsilon^2$) and $B^{12}_2=0$ (or $\bar B^{12}_2=0$).

Consider the first case. According to (\ref{l.12}),
the matrices $B^{ln}_2$ and $\bar B^{ln}_2$ are
proportional to each other. Substitution of
(\ref{l.11}) into (\ref{l.7a}) and (\ref{l.8a})
demonstrates the proportionality of vectors $A^{l}_2$
and $\bar A^{l}_2$. This means that the polynomials $R_2[{\bf z}(\lambda)]$
and $\bar R_2[{\bf
z}(\lambda)]$, should also be proportional to each other, and thus,
cannot be interpreted as Riccatians.

Let us now consider the second case. In this case the matrices
$B^{ln}_2$ and $\bar B^{ln}_2$ are diagonal. Denoting their diagonal
elements by $B^l_2$ and $\bar B^l_2$ and substituting
(\ref{l.11}) into (\ref{l.7a}) and (\ref{l.8a}), we find that
$A^{l}_2=B^l_2$ and $\bar A^{l}_2=\bar B^l_2$. This gives
\begin{eqnarray}
R_2[{\bf z}(\lambda)]=A^1_2\{\partial z_1(\lambda) + 2z_1^2(\lambda)\}+
A^2_2\{\partial z_2(\lambda) + 2z_2^2(\lambda)\}
\label{l.14}\ea\label{l.14a}\\
\bar R_2[{\bf z}(\lambda)]=\bar A^1_2\{\partial z_1(\lambda)
+ 2z_1^2(\lambda)\}+
\bar A^2_2\{\partial z_2(\lambda) + 2z_2^2(\lambda)\}
\eb\label{l.14b}
\end{eqnarray}
We see that the Riccati operator defined by relations (\ref{l.14})
is reducible, and therefore, the polynomials $R_2[{\bf z}(\lambda)]$ and
$\bar R_2[{\bf z}(\lambda)]$ cannot be considered as Riccatians.

Summarizing, we can assert that there are no simple
Riccati operators of age $|2,2\rangle$.

\subsection{Riccati operators of age $|2,3\rangle$}

Consider a general Riccati operator of age $|2,3\rangle$ determined
by formulas
\begin{eqnarray}
R_2[z(\lambda)]&=&A^i_2\partial z_i(\lambda) +
B^{ik}_2z_i(\lambda)z_k(\lambda),
\label{m.1}\ea\label{m.1a}\\
R_3[z(\lambda)]&=&A^i_3\partial^2 z_i(\lambda) +
B^{i,k}_3\partial z_i(\lambda)z_k(\lambda)+
C^{ikl}_3z_i(\lambda)z_k(\lambda)z_l(\lambda),\quad\quad
\eb\label{m.1b}
\end{eqnarray}
and try to find conditions for its coefficients under
which the polynomials $R_2[z(\lambda)]$ and $R_3[z(\lambda)]$
become Riccatians. As before, the
indices $i, k$ and $l$ in (\ref{m.1}) take the values $1$ and $2$,
and the summation over repeated indices is assumed.

The residues of polynomials (\ref{m.1a}) and (\ref{m.1b})
have the form
\begin{eqnarray}
R_{2,0}[y(\xi)]&=&-A^i_2v _i+B^{ik}_2v _iv _k,
\label{m.4}\ea\label{m.4a}\\
R_{2,1}[y(\xi)]&=&2B^{ik}_2v _iy_k(\xi),
\eb\label{m.4b}
\end{eqnarray}
and
\begin{eqnarray}
R_{3,0}[y(\xi)]&=&2A^i_3v _i-B^{i,k}_3v _iv _k+
C^{ikl}_3v _iv _kv _l,
\label{m.5}\ea\label{m.5a}\\
R_{3,1}[y(\xi)]&=&(-B^{i,l}_3v _i+
3C^{ikl}_3v _iv _k)y_l(\xi),
\eb\label{m.5b}\\
R_{3,2}[y(\xi)]&=&(B^{l,i}v _i-B^{i,l}_3v _i+
3C^{ikl}_3v _iv _k)\dot y_l(\xi)
+3C^{ikl}_3v _i y_k(\xi) y_l(\xi).\quad\quad
\ec\label{m.5c}
\end{eqnarray}
The conditions for all these residues to have a common
divisor of first degree
\begin{eqnarray}
Q[y(\xi)]=Q^ly_l(\xi)
\label{m.6}
\end{eqnarray}
are
\begin{eqnarray}
&-A^i_2v _i+B^{ik}_2v _iv _k=0,&
\label{m.7}\ea\label{m.7a}\\
&B^{li}_2v _i=e_2Q^l,&
\eb\label{m.7b}
\end{eqnarray}
for the first polynomial, and
\begin{eqnarray}
&2A^i_3v _i-B^{i,k}_3v _iv _k+
C^{ikl}_3v _iv _kv _l=0,&
\label{m.8}\ea\label{m.8a}\\
&B^{i,l}_3v _i-
3C^{ikl}_3v _iv _k=e_3Q^l,&
\eb\label{m.8b}\\
&(B^{l,i}_3-B^{i,l}_3)v _i+
3C^{ikl}_3v _iv _k=0,&
\ec\label{m.8c}\\
&3C^{ikl}_3v _i =f^k_3Q^l+f^l_3Q^k,&
\ed\label{m.8d}
\end{eqnarray}
for the second polynomial. Here $e_2$ and $e_3$ are
some unknown numbers and $f^i_3$ is some unknown
vector.  Below we shall assume that all these quantities
differ from zero.

We start with the second system.
Multiplying (\ref{m.8c}) by $v _l$ we obtain the condition
\begin{eqnarray}
(B^{l,i}_3-B^{i,l}_3)v _iv _l+
3C^{ikl}_3v _iv _kv _l=0,
\label{m.9}
\end{eqnarray}
the comparizon of which with (\ref{m.8a}) gives
\begin{eqnarray}
-A^i_3v _i+B^{ik}_3v _iv _k=0.
\label{m.10}
\end{eqnarray}
This equation is similar to equation (\ref{m.7a}) from the
first system. Comparing (\ref{m.8b}) with (\ref{m.8c})
we get another equation
\begin{eqnarray}
B^{l,i}_3v _i=e_3Q^l
\label{m.11}
\end{eqnarray}
which is obviously similar to equation (\ref{m.7b}). From
(\ref{m.7b}) and (\ref{m.11}) it follows that
\begin{eqnarray}
\frac{1}{e_2}B^{li}_2v _i=\frac{1}{e_3}B^{l,i}_3v _i
\label{m.12}
\end{eqnarray}
or, equivalently, in the matrix form,
\begin{eqnarray}
B_3 B_2^{-1}{\bf v} =\epsilon{\bf v} .
\label{m.13}
\end{eqnarray}
The last condition means that the vectors ${\bf v}$ are the
eigenvectors of the matrix $B_3 B_2^{-1}$, while
the fractions $\epsilon=e_3 e_2^{-1}$ are its eigenvalues.

The next step is to find the vectors $f^i_3$.
Multiplying (\ref{m.8d}) by $v _kv _l$ and using (\ref{m.9})
we obtain
\begin{eqnarray}
(f^k_3v _k)(Q^lv _l)=0
\label{m.14}
\end{eqnarray}
which gives
\begin{eqnarray}
(f^k_3v _k)=0
\label{m.15}
\end{eqnarray}
provided that the second factor in (\ref{m.14}) differs
from zero. Now multiplying (\ref{m.8d}) by $v _k$ and
using (\ref{m.15}) we get the condition
\begin{eqnarray}
3C^{ikl}_3v _iv _k =f^l_3(Q^kv _k),
\label{m.16}
\end{eqnarray}
which together with another condition
\begin{eqnarray}
B^{i,k}_3v _iv _k=e_3(Q^kv _k),
\label{m.17}
\end{eqnarray}
obtained by multiplying (\ref{m.11}) by $v _k$, enables
one to determine $f^l_3$:
\begin{eqnarray}
f^l_3=e_3
\frac{3C^{ikl}_3v _iv _k}{B^{i,k}_3v _iv _k}.
\label{m.18}
\end{eqnarray}
After using (\ref{m.8c}) we get
\begin{eqnarray}
f^l_3=e_3
\frac{(B^{l,i}_3-B^{i,l}_3)v _i}{B^{i,k}_3v _iv _k}.
\label{m.19}
\end{eqnarray}
Substitution of (\ref{m.19}) into (\ref{m.8d}) gives
\begin{eqnarray}
3C^{ikl}_3v _i=(e_3Q^k)
\frac{(B^{l,i}_3-B^{i,l}_3)v _i}{B^{i,j}_3v _iv _j}+
(e_3Q^l)
\frac{(B^{k,i}_3-B^{i,k}_3)v _i}{B^{i,j}_3v _iv _j},
\label{m.20}
\end{eqnarray}
or, after taking into account (\ref{m.11}),
\begin{eqnarray}
3C^{ikl}_3v _i=
\frac{(B^{i,k}_3B^{l,j}_3+B^{i,l}_3B^{k,j}_3-
2B^{k,i}_3B^{l,j}_3)v _iv _j}{B^{i,j}_3v _iv _j}.
\label{m.21}
\end{eqnarray}
This formula will play the central role in our further considerations.

We know that $B_3$ and $B_2$ are $2\times 2$ matrices, and
therefore the spectral problem (\ref{m.13}) should have two
linearly independent eigenvectors. Denote these
eigenvectors by ${\bf v}^n=\{v^n_i\}$ and note that
the basis in the space to which they belong,
can always be chosen in such a way as to guarantee their
coincidence with orthonormal repers
\begin{eqnarray}
v _i^n=\delta_i^n.
\label{m.22}
\end{eqnarray}
Substituting (\ref{m.22})
into (\ref{m.21}) we then obtain
\begin{eqnarray}
3C^{nkl}_3=
\frac{B^{n,k}_3B^{l,n}_3+B^{n,l}_3B^{k,n}_3-
2B^{k,n}_3B^{l,n}_3}{B^{n,n}_3}.
\label{m.23}
\end{eqnarray}
Let $\epsilon^n$ denote the eigenvalues
of the matrix
$B_3 B_2^{-1}$ corresponding to the
orthonormalized eigenvectors ${\bf v}^n$. Then from
(\ref{m.22}) and (\ref{m.12}) it follows that
\begin{eqnarray}
B^{l,n}_3=\epsilon^nB^{ln}_2.
\label{m.24}
\end{eqnarray}
Substituting (\ref{m.24}) into (\ref{m.23}) and taking into
account the fact that the matrix $B_2$ is symmetric, we
obtain
\begin{eqnarray}
3C^{nkl}_3=(\epsilon^n+\epsilon^l-2\epsilon^k)
\frac{B^{kn}_2B^{kl}_2}{B^{kk}_2}.
\label{m.25}
\end{eqnarray}
Remember now that the tensor $3C^{nkl}_3$ is totally
symmetric. At the same time, the expression in the right
hand side of (\ref{m.25}) is automatically symmetric only
with respect to indices $n$ and $l$. In order to guarantee
the total symmetry of this expression it is sufficient to
require its symmetry with respect to indices $l$ and $k$.
This gives the condition
\begin{eqnarray}
(\epsilon^n+\epsilon^l-2\epsilon^k) \frac{B^{kn}_2}{B^{kk}_2}B^{kl}_2=
(\epsilon^n+\epsilon^k-2\epsilon^l) \frac{B^{ln}_2}{B^{ll}_2}B^{kl}_2.
\label{m.26}
\end{eqnarray}
Taking $k=n$
in  (\ref{m.26}) we obtain
\begin{eqnarray}
(\epsilon^l-\epsilon^n)B^{ln}\left(1+ 2\frac{B^{ln}_2}{B^{ll}_2}\right)=0.
\label{m.27}
\end{eqnarray}

Now note that the relation (\ref{m.27}) can be realized in
three cases: 1) if $\epsilon^1=\epsilon^2$ and  $B^{ln}_2$ is arbitrary,
2) if $\epsilon^1\neq\epsilon^2$ and $B^{ln}_2$ is diagonal, and
3) if $\epsilon^1\neq\epsilon^2$ and
$(1+ 2B^{ln}_2/B^{ll}_2)=0$. Repeating the reasonings of
the previous subsection it is not difficult to show that
the first possibility leads to a system of dependent
Riccati polynomials which cannot be interpreted as
Riccatians, and the second possibility leads to a reducible
system of Riccati polynomials which also cannot be
interpreted as Riccatians.

Consider the third case. In this case we have
\begin{eqnarray}
B^{ln}_2=-\frac{1}{2}B^{ll}_2.
\label{m.28a}
\end{eqnarray}
Because of the symmetry of matrix $B_2$ we also have
\begin{eqnarray}
B^{ln}_2=-\frac{1}{2}B^{nn}_2,
\label{m.28b}
\end{eqnarray}
which means that
\begin{eqnarray}
B^{11}_2=B^{22}_2=B, \quad B^{12}_2=B^{21}_2=-\frac{B}{2}.
\label{m.32}
\end{eqnarray}
Substituting (\ref{m.22}) into (\ref{m.7a}) we find
\begin{eqnarray}
A^{1}_2=B, \quad A^{2}_2=B.
\label{m.33}
\end{eqnarray}
In order to compute the matrix $B^{i,k}_3$ by means of
formula (\ref{m.24}) we need the eigenvalues $e^1$ and $e^2$.
However, it is easily seen that there are no formulas
imposing some constraints on $e^1$ and $e^2$. This means
that these eigenvalues can be considered as {\it free parameters}.
For the sake of further convenience, we introduce instead
of $e^1$ and $e^2$ two other parameters, $C$ and $D$, by formulas
\begin{eqnarray}
e^1=2\frac{C+D}{B} \quad e^2=2\frac{C-D}{B}.
\label{m.34}
\end{eqnarray}
Then, using (\ref{m.34}), (\ref{m.24}) and (\ref{m.32}) we get
\begin{eqnarray}
B^{1,1}_3=2C+2D, \ \ B^{2,2}_3=2C-2D, \ \
B^{1,2}_3=-C+D, \ \ B^{2,1}_3=-C-D.
\label{m.35}
\end{eqnarray}
Formula (\ref{m.10}) after applying to it (\ref{m.22}) gives
\begin{eqnarray}
A^{1}_3=C+D, \quad A^{2}_3=C-D.
\label{m.36}
\end{eqnarray}
Finally, using (\ref{m.25}) we obtain:
\begin{eqnarray}
C^{111}_3=0, \quad C^{222}_3=0, \nonumber\\
C^{112}_3=C^{121}_3=C^{211}_3=2D, \nonumber\\
C^{221}_3=C^{212}_3=C^{122}_3=-2D.
\label{m.37}
\end{eqnarray}

Collecting the obtained expressions for the coefficients
of Riccati polynomials and writing down final and most general
expressions for Riccatians (\ref{m.1a}) and (\ref{m.1b}),
we can see that the polynomial proportional to $C$
is nothing else than the derivative of $R_2[z(\lambda)]$.
Thus, without loss of generality, we can take $C=0$.
Choosing the remaining normalization coefficients as $B=1$
and $D=1$, we obtain final expression for the system of two
Riccatians of orders $2$ and $3$.
\begin{eqnarray}
R_2[z(\lambda)] &=& \partial z_1(\lambda)
+\partial z_2(\lambda)\nonumber\\
&&+z_1(\lambda)z_2(\lambda)
+z_2(\lambda)z_2(\lambda)
-z_1(\lambda)z_2(\lambda),\quad
\label{m.39}\ea\label{m.39a}\\
R_3[z(\lambda)] &=& \partial^2 z_1(\lambda)
-\partial^2 z_2(\lambda)
+2\partial z_1(\lambda)z_1(\lambda)
-2\partial z_2(\lambda)z_2(\lambda)+\nonumber\\
&&+\partial z_1(\lambda) z_2(\lambda)
-\partial z_2(\lambda) z_1(\lambda)+\nonumber\\
&&+2z_1(\lambda)z_2(\lambda)z_2(\lambda)
-2z_2(\lambda)z_2(\lambda)z_1(\lambda).
\eb\label{m.39b}
\end{eqnarray}
The regularizator of the system (\ref{m.39})
consists of two elements
\begin{eqnarray}
v ^1=(1,0), \quad Q^1[z(\lambda)]=z_1(\lambda)-\frac{1}{2}z_2(\lambda),
\nonumbe
r\\
v ^2=(0,1), \quad Q^2[z(\lambda)]=z_2(\lambda)-\frac{1}{2}z_1(\lambda).
\label{m.40}
\end{eqnarray}
These expressions immediately follow from formulas
(\ref{m.7a}) and (\ref{m.11}) after taking into account
explicit expressions for matrices $B_2, B_3$ and numbers
$e_2, e_3$.

\section{Some other examples of Riccatians.}

The calculations given in previous section can be repeated
for other Riccati operators. Being essentially the
same they, however, become technically more and more
complicated with increasing the age of Riccati
operators. For this reason we were forced to use the
programs of analytic calculations like REDUCE and MATHEMATICA.
We have analysed {\it all} Riccati operators of
ages $|n_1,\ldots, n_r\rangle$ with $r\le 3$ and $n_r\le 6$.
We present here the results of such calculations
ommiting details of their derivations. Below we give only the
list of simple Riccati operators generating the systems of Riccatians.
For the sake of completeness this list will also contain the
cases already discussed in previous section.

\subsection{Riccati operator of age $|2\rangle$}

\noindent Riccatian:
\begin{eqnarray}
R_2[{\bf z}] &=&
 \partial z_1
 + z_1^2.
\label{n.1}
\end{eqnarray}
\noindent Regularizator:
\begin{eqnarray}
{\bf v}^1=1, \quad Q^1[{\bf y}]=y.
\label{o.1}
\end{eqnarray}

\subsection{Riccati operator of age $|2,3\rangle$}
\noindent Riccatians\footnote{Here the role of Riccatian $R_3$ plays
the combination $R_3+\partial R_2$ of Riccatians (\ref{m.39}).}:
\begin{eqnarray}
R_2[{\bf z}] &=&
 \partial z_1
 + \partial z_2
 + z_1^2
 - z_1 z_2
 + z_2^2,
\label{n.2}\ea\label{n.2a}\\[1cm]
R_3[{\bf z}] &=&
 \partial^2 z_1
 + 2 \partial z_1 z_1
 - \partial z_2 z_1
 + z_1^2 z_2
 - z_1 z_2^2.
\eb\label{n.2b}
\end{eqnarray}
\noindent Regularizator:
\begin{eqnarray}
{\bf v}^1=(1,0), &&\quad Q^1[{\bf y}]=y_1-\frac{1}{2}y_2,\nonumber\\
{\bf v}^2=(0,1), &&\quad Q^2[{\bf y}]=y_2-\frac{1}{2}y_1.
\label{o.2}
\end{eqnarray}

\subsection{Riccati operator of age $|2,4\rangle$}

\noindent Riccatians:
\begin{eqnarray}
R_2[{\bf z}] &=&
 2 \partial z_1
 + \partial z_2
 + 2 z_1^2
 - 2 z_1 z_2
 + z_2^2,
\label{n.3}\ea\label{n.3a}\\[1cm]
R_4[{\bf z}] &=&
  \partial^3 z_1
 + 2 \partial^2 z_1 z_1
 - \partial^2 z_2 z_1
 + (\partial z_1)^2
 - \partial z_1 \partial z_2
\nonumber\\ &-& 2 \partial z_1 z_1^2
 + 4 \partial z_1 z_1 z_2
 - \partial z_1 z_2^2
 + \partial z_2 z_1^2
 - 2 \partial z_2 z_1 z_2
\nonumber\\ &-& z_1^4
 + 2 z_1^3 z_2
 - z_1^2 z_2^2.
\eb\label{n.3b}
\end{eqnarray}
\noindent Regularizator:
\begin{eqnarray}
{\bf v}^1=(1,0), &&\quad Q^1[{\bf y}]=2y_1-y_2,\nonumber\\
{\bf v}^2=(0,1), &&\quad Q^2[{\bf y}]=y_2-y_1.
\label{o.3}
\end{eqnarray}

\subsection{Riccati operator of age $|2,6\rangle$}

\noindent Riccatians:
\begin{eqnarray}
R_2[{\bf z}] &=&
  3 \partial z_1
 +  \partial z_2
 + 3 z_1^2
 - 3 z_1 z_2
 +  z_2^2,
\label{n.4}\ea\label{n.4a}\\[1cm]
R_6[{\bf z}] &=&
 5 \partial^5 z_1
 + \partial^5 z_2
\nonumber\\ &+& 10 \partial^4 z_1 z_1
 - 3 \partial^4 z_1 z_2
 - 5 \partial^4 z_2 z_1
 + 2 \partial^4 z_2 z_2
\nonumber\\ &+& 17 \partial^3 z_1 \partial z_1
 - 19 \partial^3 z_1 \partial z_2
 - 21 \partial^3 z_2 \partial z_1
 + 7 \partial^3 z_2 \partial z_2
\nonumber\\ &+& 10 (\partial^2 z_1)^2
 - 30 \partial^2 z_1 \partial^2 z_2
 + 5 (\partial^2 z_2)^2
 - 23 \partial^3 z_1 z_1^2
\nonumber\\ &+& 27 \partial^3 z_1 z_1 z_2
 - 7 \partial^3 z_1 z_2^2
 - \partial^3 z_2 z_1^2
 - \partial^3 z_2 z_1 z_2
\nonumber\\ &-& \partial^3 z_2 z_2^2
 - 126 \partial^2 z_1 \partial z_1 z_1
 + 63 \partial^2 z_1 \partial z_1 z_2
 + 42 \partial^2 z_1 \partial z_2 z_1
\nonumber\\ &-& 21 \partial^2 z_1 \partial z_2 z_2
 + 23 \partial^2 z_2 \partial z_1 z_1
 - 12 \partial^2 z_2 \partial z_1 z_2
 - \partial^2 z_2 \partial z_2 z_1
\nonumber\\ &-& 6 \partial^2 z_2 \partial z_2 z_2
 - 42 (\partial z_1)^3
 + 46 (\partial z_1)^2 \partial z_2
 - 10 \partial z_1 (\partial z_2)^2
\nonumber\\ &-& 2 (\partial z_2)^3
 - 46 \partial^2 z_1 z_1^3
 + 57 \partial^2 z_1 z_1^2 z_2
 - 21 \partial^2 z_1 z_1 z_2^2
\nonumber\\ &+& 3 \partial^2 z_1 z_2^3
 + 23 \partial^2 z_2 z_1^3
 - 29 \partial^2 z_2 z_1^2 z_2
 + 11 \partial^2 z_2 z_1 z_2^2
\nonumber\\ &-& 2 \partial^2 z_2 z_2^3
 - 114 (\partial z_1)^2 z_1^2
 + 90 (\partial z_1)^2 z_1 z_2
 - 17 (\partial z_1)^2 z_2^2
\nonumber\\ &+& 122 \partial z_1 \partial z_2 z_1^2
 - 98 \partial z_1 \partial z_2 z_1 z_2
 + 22 \partial z_1 \partial z_2 z_2^2
 - 28 (\partial z_2)^2 z_1^2
\nonumber\\ &+& 22 (\partial z_2)^2 z_1 z_2
 - 6 (\partial z_2)^2 z_2^2
 + 12 \partial z_1 z_1^4
 - 48 \partial z_1 z_1^3 z_2
\nonumber\\ &+& 50 \partial z_1 z_1^2 z_2^2
 - 18 \partial z_1 z_1 z_2^3
 + 2 \partial z_1 z_2^4
 - 2 \partial z_2 z_1^4
\nonumber\\ &+& 16 \partial z_2 z_1^3 z_2
 - 16 \partial z_2 z_1^2 z_2^2
 + 4 \partial z_2 z_1 z_2^3
 + 4 z_1^6
\nonumber\\ &-& 12 z_1^5 z_2
 + 13 z_1^4 z_2^2
 - 6 z_1^3 z_2^3
 + z_1^2 z_2^4.
\eb\label{n.4b}
\end{eqnarray}
\noindent Regularizator:
\begin{eqnarray}
{\bf v}^1=(1,0), &&\quad Q^1[{\bf y}]=3y_1-\frac{3}{2}y_2,\nonumber\\
{\bf v}^2=(0,1), &&\quad Q^2[{\bf y}]=y_2-\frac{3}{2}y_1.
\label{o.4}
\end{eqnarray}

\subsection{Riccati operators of age $|2,3,4\rangle$}

\noindent Riccatians:
\begin{eqnarray}
R_2[{\bf z}] &=&
 \partial z_1
 + \partial z_2
 + \partial z_3
\nonumber\\ &+& z_1^2
 - z_1 z_2
 + z_2^2
 - z_2 z_3
 + z_3^2,
\label{n.5}\ea\label{n.5a}\\[1cm]
R_3[{\bf z}] &=&
 2 \partial^2 z_1
 + \partial^2 z_2
\nonumber\\ &+& 4 \partial z_1 z_1
 - \partial z_1 z_2
 - 2 \partial z_2 z_1
 + 2 \partial z_2 z_2
 - \partial z_3 z_2
\nonumber\\ &+& z_1^2 z_2
 - z_1 z_2^2
 + z_2^2 z_3
 - z_2 z_3^2,
\eb\label{n.5b}\\[1cm]
R_4[{\bf z}] &=&
 \partial^3 z_1
 + 2 \partial^2 z_1 z_1
 - \partial^2 z_2 z_1
 + 2 (\partial z_1)^2
 - \partial z_1 \partial z_2
\nonumber\\ &-& \partial z_1 \partial z_3
 + 2 \partial z_1 z_1 z_2
 - \partial z_1 z_2^2
 + \partial z_1 z_2 z_3
 - \partial z_1 z_3^2
\nonumber\\ &+& \partial z_2 z_1^2
 - 2 \partial z_2 z_1 z_2
 - \partial z_3 z_1^2
 + \partial z_3 z_1 z_2
 + z_1^2 z_2 z_3
\nonumber\\ &-& z_1^2 z_3^2
 - z_1 z_2^2 z_3
 + z_1 z_2 z_3^2.
\ec\label{n.5c}
\end{eqnarray}
\noindent Regularizator:
\begin{eqnarray}
{\bf v}^1=(1,0,0), &&\quad Q^1[{\bf y}]=y_1-\frac{1}{2}y_2,\nonumber\\
{\bf v}^2=(0,1,0), &&\quad Q^2[{\bf y}]=y_2-\frac{1}{2}y_1-\frac{1}{2}y_3,
\nonumber\\
{\bf v}^3=(0,0,1), &&\quad Q^3[{\bf y}]=y_3-\frac{1}{2}y_2.
\label{o.5}
\end{eqnarray}

\subsection{Riccati operator of age $|2,4,6\rangle$. First case}

\noindent Riccatians:
\begin{eqnarray}
R_2[{\bf z}] &=&
 2 \partial z_1
 + 2 \partial z_2
 + \partial z_3
\nonumber\\ &+& 2 z_1^2
 - 2 z_1 z_2
 + 2 z_2^2
 - 2 z_2 z_3
 + z_3^2,
\label{n.6}\ea\label{n.6a}\\[1cm]
R_4[{\bf z}] &=&
 6 \partial^3 z_1
 + 3 \partial^3 z_2
 + \partial^3 z_3
 + 12 \partial^2 z_1 z_1
 - 3 \partial^2 z_1 z_2
\nonumber\\ &-& 6 \partial^2 z_2 z_1
 + 6 \partial^2 z_2 z_2
 - 2 \partial^2 z_2 z_3
 - 3 \partial^2 z_3 z_2
 + 2 \partial^2 z_3 z_3
\nonumber\\ &+& 11 (\partial z_1)^2
 - 11 \partial z_1 \partial z_2
 - 2 \partial z_1 \partial z_3
 + 5 (\partial z_2)^2
 - 5 \partial z_2 \partial z_3
\nonumber\\ &+& 2 (\partial z_3)^2
 - 2 \partial z_1 z_1^2
 + 8 \partial z_1 z_1 z_2
 - 5 \partial z_1 z_2^2
 + 4 \partial z_1 z_2 z_3
\nonumber\\ &-& 2 \partial z_1 z_3^2
 + \partial z_2 z_1^2
 - 4 \partial z_2 z_1 z_2
 - 2 \partial z_2 z_2^2
 + 4 \partial z_2 z_2 z_3
\nonumber\\ &-& \partial z_2 z_3^2
 - 2 \partial z_3 z_1^2
 + 2 \partial z_3 z_1 z_2
 + \partial z_3 z_2^2
 - 2 \partial z_3 z_2 z_3
\nonumber\\ &-& z_1^4
 + 2 z_1^3 z_2
 - 3 z_1^2 z_2^2
 + 4 z_1^2 z_2 z_3
 - 2 z_1^2 z_3^2
 + 2 z_1 z_2^3
\nonumber\\ &-& 4 z_1 z_2^2 z_3
 + 2 z_1 z_2 z_3^2
 - z_2^4
 + 2 z_2^3 z_3
 - z_2^2 z_3^2,
\eb\label{n.6b}\\[1cm]
R_6[{\bf z}] &=&
 \partial^5 z_1
 + 2 \partial^4 z_1 z_1
 - \partial^4 z_2 z_1
 + 7 \partial^3 z_1 \partial z_1
 - 2 \partial^3 z_1 \partial z_2
\nonumber\\ &-& \partial^3 z_1 \partial z_3
 - 3 \partial^3 z_2 \partial z_1
 - \partial^3 z_3 \partial z_1
 + 6 (\partial^2 z_1)^2
 - 4 \partial^2 z_1 \partial^2 z_2
\nonumber\\ &-& 2 \partial^2 z_1 \partial^2 z_3
 - \partial^3 z_1 z_1^2
 + 3 \partial^3 z_1 z_1 z_2
 - 2 \partial^3 z_1 z_2^2
 + 2 \partial^3 z_1 z_2 z_3
\nonumber\\ &-& \partial^3 z_1 z_3^2
 + \partial^3 z_2 z_1^2
 - 2 \partial^3 z_2 z_1 z_2
 - \partial^3 z_3 z_1^2
 + \partial^3 z_3 z_1 z_2
\nonumber\\ &-& 2 \partial^2 z_1 \partial z_1 z_1
 + 7 \partial^2 z_1 \partial z_1 z_2
 + 4 \partial^2 z_1 \partial z_2 z_1
 - 8 \partial^2 z_1 \partial z_2 z_2
\nonumber\\ &+& 4 \partial^2 z_1 \partial z_2 z_3
 - 2 \partial^2 z_1 \partial z_3 z_1
 + 4 \partial^2 z_1 \partial z_3 z_2
 - 4 \partial^2 z_1 \partial z_3 z_3
\nonumber\\ &+& 5 \partial^2 z_2 \partial z_1 z_1
 - 6 \partial^2 z_2 \partial z_1 z_2
 + 2 \partial^2 z_2 \partial z_1 z_3
 - 4 \partial^2 z_2 \partial z_2 z_1
\nonumber\\ &+& \partial^2 z_2 \partial z_3 z_1
 - 4 \partial^2 z_3 \partial z_1 z_1
 + 3 \partial^2 z_3 \partial z_1 z_2
 - 2 \partial^2 z_3 \partial z_1 z_3
\nonumber\\ &+& 2 \partial^2 z_3 \partial z_2 z_1
 - 2 (\partial z_1)^3
 + 5 (\partial z_1)^2 \partial z_2
 - (\partial z_1)^2 \partial z_3
\nonumber\\ &-& 5 \partial z_1 (\partial z_2)^2
 + 5 \partial z_1 \partial z_2 \partial z_3
 - 2 \partial z_1 (\partial z_3)^2
 - 2 \partial^2 z_1 z_1^3
\nonumber\\ &+& 3 \partial^2 z_1 z_1^2 z_2
 - 3 \partial^2 z_1 z_1 z_2^2
 + 4 \partial^2 z_1 z_1 z_2 z_3
 - 2 \partial^2 z_1 z_1 z_3^2
\nonumber\\ &+& \partial^2 z_2 z_1^3
 - 2 \partial^2 z_2 z_1^2 z_2
 + 2 \partial^2 z_2 z_1^2 z_3
 + 2 \partial^2 z_2 z_1 z_2^2
\nonumber\\ &-& 4 \partial^2 z_2 z_1 z_2 z_3
 + \partial^2 z_2 z_1 z_3^2
 + \partial^2 z_3 z_1^2 z_2
 - 2 \partial^2 z_3 z_1^2 z_3
\nonumber\\ &-& \partial^2 z_3 z_1 z_2^2
 + 2 \partial^2 z_3 z_1 z_2 z_3
 - 2 (\partial z_1)^2 z_1^2
 + 2 (\partial z_1)^2 z_1 z_2
\nonumber\\ &-& (\partial z_1)^2 z_2^2
 + 2 (\partial z_1)^2 z_2 z_3
 - (\partial z_1)^2 z_3^2
 - 6 \partial z_1 \partial z_2 z_1 z_2
\nonumber\\ &+& 8 \partial z_1 \partial z_2 z_1 z_3
 + 2 \partial z_1 \partial z_2 z_2^2
 - 4 \partial z_1 \partial z_2 z_2 z_3
 + \partial z_1 \partial z_2 z_3^2
\nonumber\\ &+& 2 \partial z_1 \partial z_3 z_1^2
 + 4 \partial z_1 \partial z_3 z_1 z_2
 - 8 \partial z_1 \partial z_3 z_1 z_3
 - \partial z_1 \partial z_3 z_2^2
\nonumber\\ &+& 2 \partial z_1 \partial z_3 z_2 z_3
 - (\partial z_2)^2 z_1^2
 + 4 (\partial z_2)^2 z_1 z_2
 - 4 (\partial z_2)^2 z_1 z_3
\nonumber\\ &+& 3 \partial z_2 \partial z_3 z_1^2
 - 6 \partial z_2 \partial z_3 z_1 z_2
 + 4 \partial z_2 \partial z_3 z_1 z_3
 - 2 (\partial z_3)^2 z_1^2
\nonumber\\ &+& 2 (\partial z_3)^2 z_1 z_2
 + 2 \partial z_1 z_1^2 z_2^2
 - 4 \partial z_1 z_1^2 z_2 z_3
 + 2 \partial z_1 z_1^2 z_3^2
\nonumber\\ &-& 4 \partial z_1 z_1 z_2^3
 + 8 \partial z_1 z_1 z_2^2 z_3
 - 4 \partial z_1 z_1 z_2 z_3^2
 + \partial z_1 z_2^4
\nonumber\\ &-& 2 \partial z_1 z_2^3 z_3
 + \partial z_1 z_2^2 z_3^2
 - \partial z_2 z_1^4
 + 2 \partial z_2 z_1^3 z_2
\nonumber\\ &-& 4 \partial z_2 z_1^2 z_2^2
 + 4 \partial z_2 z_1^2 z_2 z_3
 - \partial z_2 z_1^2 z_3^2
 + 4 \partial z_2 z_1 z_2^3
\nonumber\\ &-& 6 \partial z_2 z_1 z_2^2 z_3
 + 2 \partial z_2 z_1 z_2 z_3^2
 + \partial z_3 z_1^4
 - 2 \partial z_3 z_1^3 z_2
\nonumber\\ &+& 3 \partial z_3 z_1^2 z_2^2
 - 2 \partial z_3 z_1^2 z_2 z_3
 - 2 \partial z_3 z_1 z_2^3
 + 2 \partial z_3 z_1 z_2^2 z_3
\nonumber\\ &+& z_1^4 z_2^2
 - 2 z_1^4 z_2 z_3
 + z_1^4 z_3^2
 - 2 z_1^3 z_2^3
\nonumber\\ &+& 4 z_1^3 z_2^2 z_3
 - 2 z_1^3 z_2 z_3^2
 + z_1^2 z_2^4
 - 2 z_1^2 z_2^3 z_3
 + z_1^2 z_2^2 z_3^2.
\ec\label{n.6c}
\end{eqnarray}
\noindent Regularizator:
\begin{eqnarray}
{\bf v}^1=(1,0,0), &&\quad Q^1[{\bf y}]=2y_1-y_2,\nonumber\\
{\bf v}^2=(0,1,0), &&\quad Q^2[{\bf y}]=2y_2-y_1-2y_3,
\nonumber\\
{\bf v}^3=(0,0,1), &&\quad Q^3[{\bf y}]=y_3-y_2.
\label{o.6}
\end{eqnarray}

\subsection{Riccati operators of age $|2,4,6\rangle$.
Second case}

\noindent Riccatians:
\begin{eqnarray}
R_2[{\bf z}] &=&
  \partial z_1
 + \partial z_2
 + 2 \partial z_3
\nonumber\\ &+& 2 z_1^2
 -  z_1 z_2
 +  z_2^2
 - 2 z_2 z_3
 + 2 z_3^2,
\label{n.7}\ea\label{n.7a}\\[1cm]
R_4[{\bf z}] &=&
 10 \partial^3 z_1
 + 6 \partial^3 z_2
 + 8 \partial^3 z_3
 + 20 \partial^2 z_1 z_1
 - 6 \partial^2 z_1 z_2
\nonumber\\ &-& 10 \partial^2 z_2 z_1
 + 12 \partial^2 z_2 z_2
 - 8 \partial^2 z_2 z_3
 - 12 \partial^2 z_3 z_2
 + 16 \partial^2 z_3 z_3
\nonumber\\ &+& 19 (\partial z_1)^2
 - 18 \partial z_1 \partial z_2
 - 8 \partial z_1 \partial z_3
 + 11 (\partial z_2)^2
 - 24 \partial z_2 \partial z_3
\nonumber\\ &+& 16 (\partial z_3)^2
 - 2 \partial z_1 z_1^2
 + 10 \partial z_1 z_1 z_2
 - 6 \partial z_1 z_2^2
 + 8 \partial z_1 z_2 z_3
\nonumber\\ &-& 8 \partial z_1 z_3^2
 + 2 \partial z_2 z_1^2
 - 6 \partial z_2 z_1 z_2
 - 2 \partial z_2 z_2^2
 + 12 \partial z_2 z_2 z_3
\nonumber\\ &-& 8 \partial z_2 z_3^2
 - 8 \partial z_3 z_1^2
 + 8 \partial z_3 z_1 z_2
 - 8 \partial z_3 z_2 z_3
 - z_1^4
\nonumber\\ &+& 2 z_1^3 z_2
 - 3 z_1^2 z_2^2
 + 8 z_1^2 z_2 z_3
 - 8 z_1^2 z_3^2
 + 2 z_1 z_2^3
\nonumber\\ &-& 8 z_1 z_2^2 z_3
 + 8 z_1 z_2 z_3^2
 - z_2^4
 + 4 z_2^3 z_3
 - 4 z_2^2 z_3^2,
\eb\label{n.7b}\\[1cm]
R_6[{\bf z}] &=&
 5 \partial^5 z_1
 + \partial^5 z_2
 + 10 \partial^4 z_1 z_1
 - \partial^4 z_1 z_2
 - 5 \partial^4 z_2 z_1
\nonumber\\ &+& 2 \partial^4 z_2 z_2
 - 2 \partial^4 z_3 z_2
 + 37 \partial^3 z_1 \partial z_1
 - 11 \partial^3 z_1 \partial z_2
\nonumber\\ &-& 12 \partial^3 z_1 \partial z_3
 - 17 \partial^3 z_2 \partial z_1
 + 7 \partial^3 z_2 \partial z_2
 - 4 \partial^3 z_2 \partial z_3
\nonumber\\ &-& 8 \partial^3 z_3 \partial z_1
 - 8 \partial^3 z_3 \partial z_2
 + 30 (\partial^2 z_1)^2
 - 20 \partial^2 z_1 \partial^2 z_2
\nonumber\\ &-& 20 \partial^2 z_1 \partial^2 z_3
 + 5 (\partial^2 z_2)^2
 - 10 \partial^2 z_2 \partial^2 z_3
 - 3 \partial^3 z_1 z_1^2
\nonumber\\ &+& 11 \partial^3 z_1 z_1 z_2
 - 7 \partial^3 z_1 z_2^2
 + 12 \partial^3 z_1 z_2 z_3
 - 12 \partial^3 z_1 z_3^2
\nonumber\\ &+& 3 \partial^3 z_2 z_1^2
 - 7 \partial^3 z_2 z_1 z_2
 - \partial^3 z_2 z_2^2
 + 6 \partial^3 z_2 z_2 z_3
\nonumber\\ &-& 4 \partial^3 z_2 z_3^2
 - 8 \partial^3 z_3 z_1^2
 + 8 \partial^3 z_3 z_1 z_2
 - 4 \partial^3 z_3 z_2 z_3
\nonumber\\ &-& 6 \partial^2 z_1 \partial z_1 z_1
 + 27 \partial^2 z_1 \partial z_1 z_2
 + 18 \partial^2 z_1 \partial z_2 z_1
 - 27 \partial^2 z_1 \partial z_2 z_2
\nonumber\\ &+& 20 \partial^2 z_1 \partial z_2 z_3
 - 24 \partial^2 z_1 \partial z_3 z_1
 + 24 \partial^2 z_1 \partial z_3 z_2
 - 40 \partial^2 z_1 \partial z_3 z_3
\nonumber\\ &+& 23 \partial^2 z_2 \partial z_1 z_1
 - 24 \partial^2 z_2 \partial z_1 z_2
 + 8 \partial^2 z_2 \partial z_1 z_3
 - 19 \partial^2 z_2 \partial z_2 z_1
\nonumber\\ &-& 6 \partial^2 z_2 \partial z_2 z_2
 + 18 \partial^2 z_2 \partial z_2 z_3
 + 12 \partial^2 z_2 \partial z_3 z_1
 + 12 \partial^2 z_2 \partial z_3 z_2
\nonumber\\ &-& 20 \partial^2 z_2 \partial z_3 z_3
 - 40 \partial^2 z_3 \partial z_1 z_1
 + 24 \partial^2 z_3 \partial z_1 z_2
 - 16 \partial^2 z_3 \partial z_1 z_3
\nonumber\\ &+& 20 \partial^2 z_3 \partial z_2 z_1
 + 6 \partial^2 z_3 \partial z_2 z_2
 - 16 \partial^2 z_3 \partial z_2 z_3
 - 12 \partial^2 z_3 \partial z_3 z_2
\nonumber\\ &-& 6 (\partial z_1)^3
 + 22 (\partial z_1)^2 \partial z_2
 - 20 (\partial z_1)^2 \partial z_3
 - 22 \partial z_1 (\partial z_2)^2
\nonumber\\ &+& 40 \partial z_1 \partial z_2 \partial z_3
 - 16 \partial z_1 (\partial z_3)^2
 - 2 (\partial z_2)^3
 + 12 (\partial z_2)^2 \partial z_3
\nonumber\\ &-& 16 \partial z_2 (\partial z_3)^2
 - 6 \partial^2 z_1 z_1^3
 + 9 \partial^2 z_1 z_1^2 z_2
 - 9 \partial^2 z_1 z_1 z_2^2
\nonumber\\ &+& 24 \partial^2 z_1 z_1 z_2 z_3
 - 24 \partial^2 z_1 z_1 z_3^2
 + \partial^2 z_1 z_2^3
 - 4 \partial^2 z_1 z_2^2 z_3
\nonumber\\ &+& 4 \partial^2 z_1 z_2 z_3^2
 + 3 \partial^2 z_2 z_1^3
 - 5 \partial^2 z_2 z_1^2 z_2
 + 8 \partial^2 z_2 z_1^2 z_3
\nonumber\\ &+& 5 \partial^2 z_2 z_1 z_2^2
 - 20 \partial^2 z_2 z_1 z_2 z_3
 + 12 \partial^2 z_2 z_1 z_3^2
 - 2 \partial^2 z_2 z_2^3
\nonumber\\ &+& 6 \partial^2 z_2 z_2^2 z_3
 - 4 \partial^2 z_2 z_2 z_3^2
 + 4 \partial^2 z_3 z_1^2 z_2
 - 16 \partial^2 z_3 z_1^2 z_3
\nonumber\\ &-& 4 \partial^2 z_3 z_1 z_2^2
 + 16 \partial^2 z_3 z_1 z_2 z_3
 + 2 \partial^2 z_3 z_2^3
 - 4 \partial^2 z_3 z_2^2 z_3
\nonumber\\ &-& 6 (\partial z_1)^2 z_1^2
 + 6 (\partial z_1)^2 z_1 z_2
 - 5 (\partial z_1)^2 z_2^2
 + 20 (\partial z_1)^2 z_2 z_3
\nonumber\\ &-& 20 (\partial z_1)^2 z_3^2
 + 2 \partial z_1 \partial z_2 z_1^2
 - 14 \partial z_1 \partial z_2 z_1 z_2
 + 40 \partial z_1 \partial z_2 z_1 z_3
\nonumber\\ &+& 10 \partial z_1 \partial z_2 z_2^2
 - 40 \partial z_1 \partial z_2 z_2 z_3
 + 24 \partial z_1 \partial z_2 z_3^2
 + 8 \partial z_1 \partial z_3 z_1^2
\nonumber\\ &+& 16 \partial z_1 \partial z_3 z_1 z_2
 - 80 \partial z_1 \partial z_3 z_1 z_3
 - 8 \partial z_1 \partial z_3 z_2^2
 + 32 \partial z_1 \partial z_3 z_2 z_3
\nonumber\\ &-& 4 (\partial z_2)^2 z_1^2
 + 10 (\partial z_2)^2 z_1 z_2
 - 20 (\partial z_2)^2 z_1 z_3
 - 6 (\partial z_2)^2 z_2^2
\nonumber\\ &+& 12 (\partial z_2)^2 z_2 z_3
 - 4 (\partial z_2)^2 z_3^2
 + 16 \partial z_2 \partial z_3 z_1^2
 - 28 \partial z_2 \partial z_3 z_1 z_2
\nonumber\\ &+& 40 \partial z_2 \partial z_3 z_1 z_3
 + 12 \partial z_2 \partial z_3 z_2^2
 - 16 \partial z_2 \partial z_3 z_2 z_3
 - 16 (\partial z_3)^2 z_1^2
\nonumber\\ &+& 16 (\partial z_3)^2 z_1 z_2
 - 4 (\partial z_3)^2 z_2^2
 + 2 \partial z_1 z_1^2 z_2^2
 - 8 \partial z_1 z_1^2 z_2 z_3
\nonumber\\ &+& 8 \partial z_1 z_1^2 z_3^2
 - 6 \partial z_1 z_1 z_2^3
 + 24 \partial z_1 z_1 z_2^2 z_3
 - 24 \partial z_1 z_1 z_2 z_3^2
\nonumber\\ &+& 2 \partial z_1 z_2^4
 - 8 \partial z_1 z_2^3 z_3
 + 8 \partial z_1 z_2^2 z_3^2
 - 2 \partial z_2 z_1^4
\nonumber\\ &+& 4 \partial z_2 z_1^3 z_2
 - 4 \partial z_2 z_1^2 z_2^2
 + 4 \partial z_2 z_1^2 z_2 z_3
 + 4 \partial z_2 z_1 z_2^3
\nonumber\\ &-& 12 \partial z_2 z_1 z_2^2 z_3
 + 8 \partial z_2 z_1 z_2 z_3^2
 + 4 \partial z_3 z_1^4
 - 8 \partial z_3 z_1^3 z_2
\nonumber\\ &+& 8 \partial z_3 z_1^2 z_2^2
 - 8 \partial z_3 z_1^2 z_2 z_3
 - 4 \partial z_3 z_1 z_2^3
 + 8 \partial z_3 z_1 z_2^2 z_3
\nonumber\\ &+& z_1^4 z_2^2
 - 4 z_1^4 z_2 z_3
 + 4 z_1^4 z_3^2
 - 2 z_1^3 z_2^3
\nonumber\\ &+& 8 z_1^3 z_2^2 z_3
 - 8 z_1^3 z_2 z_3^2
 + z_1^2 z_2^4
 - 4 z_1^2 z_2^3 z_3
 + 4 z_1^2 z_2^2 z_3^2.
\ec\label{n.7c}
\end{eqnarray}
\noindent Regularizator:
\begin{eqnarray}
{\bf v}^1=(1,0,0), &&\quad Q^1[{\bf y}]=y_1-\frac{1}{2}y_2,\nonumber\\
{\bf v}^2=(0,1,0), &&\quad Q^2[{\bf y}]=y_2-\frac{1}{2}y_1-y_3,
\nonumber\\
{\bf v}^3=(0,0,1), &&\quad Q^3[{\bf y}]=2y_3-y_2.
\label{o.7}
\end{eqnarray}

\section{ Riccatians and simple Lie algebras}

Let us try to explane why the list of Riccatians
given in the previous section is so
interesting to us. First reason is
associated with the fact that this list obtained as a solution
of a typical analytic
problem (in whose formulation has not been used any notion
of symmetry), has purely Lie algebraic interpretation.  This
assertion (obtained after analyzing the results of previous
section) we can formulate in the form of the following theorem.

\medskip
{\bf Theorem 5.1.} {\it There exists a
correspondence between systems of simple Riccati operators
of genus 1 and simple Lie algebras. Any simple Riccati operator of age
$|n_1,\ldots, n_r\rangle$ with $r\le 3$ and $n_r\le 6$
is associated with some simple
Lie algebra ${\cal L}_r$ of rank $r$.  The corresponding
systems of Riccatians have the following properties:

1. The product of degrees of particular Riccatians forming the system
is equal to the dimension of the Weyl group for algebra ${\cal L}_r$.

2. The degrees of particular Riccatians coincide with
degrees of independent Casimir invariants for algebra
${\cal L}_r$. In particular, each system of Riccatians
contains Riccatian of second degree.

3. The regularizator of the system consists of $r$ pairs
$\{{\bf v}^n, Q^n[{\bf y}]\}, \quad
n=1,\ldots, r$ in which all vectors
${\bf v}^n$ are linearly independent.

4. It is possible to choose such a basis in
$r$-dimensional vector space  (in which function ${\bf z}$
takes its values), in which all the regularizing vectors
${\bf v}_n$ become orthonormal. In this basis (which we
shall call canonical) the regularizing
polynomials take the form
\begin{eqnarray}
Q^n[{\bf y}]=\sum_{m=1}^r(\pi^n,\pi^m)y_m
\label{5.9}
\end{eqnarray}
where $\pi^n, \ n=1,\ldots,r$
are simple roots of algebra ${\cal L}_r$ and $(\ , \ )$
denotes their scalar product.

5. In the canonical basis the second-order Riccatian has
the form
\begin{eqnarray}
R_2[{\bf z}]=\sum_{i=1}^r (\pi^i, \pi^i) \partial
z_i + \sum_{i,k=1}^r(\pi^i,\pi^k)z_i z_k.
\label{5.10}
\end{eqnarray}

6. The leading term of any Riccatian (i.e. term not
containing the derivatives) has the form
$g^{i_1,\ldots,i_n}z_{i_1}\cdots z_{i_n}$ where $n$ is the
degree of the Riccatian and $g^{i_1,\ldots,i_n}$ is an
invariant tensor of the rank $r$ generating the Cartan part
of $n$th order Casimir invariant for algebra ${\cal L}_r$.}

\medskip
This theorem is very interesting from purely
marthematical point of view, because it reveals an intriguing
relationship between the analytic proprties of Riccati operators
and their hidden symmetry properties.
Beyond any doubt, the fact that the structure of Casimir
invariants of simple Lie algebras is encoded in the
systems of Riccatians and can be obtained in absolutely
non-algebraic way, has a great theoretical significance
and deserves a careful study.

\medskip
{\bf Conjecture 5.1.} {\it There exist one-to one correspondence
between all systems of Riccatians of genus 1 and all simple
Lie algebras. The restriction to Lie algebras of ranks $r\le
3$ and to Riccati operators of ages $|n_1,\ldots, n_r\rangle$
with $r\le 3$ and $n_r\le 6$, used in theorem 5.1, is not necessary.}

\medskip
In fact, we already have a proof of this conjecture for algebras
$A_r, B_r, C_r$ of arbitrary rank $r$ and also for algebra $D_4$.
First three series have been analyzed by methods
differing from those used in the present paper,
while the $D_4$ case has been studied in the spirit of section
3\footnote{The paper with this proof is still in preparation.}.

Theorem 5.1 and its generalizations are not the only reason for which the
non-equivalent systems of Riccatians might be interesting to us.
The second and, in our opinion, the main reason follows
from the following theorem.

\medskip
{\bf Theorem 5.2.} {\it Let $R_{n_1}[z_1(\lambda),\ldots,z_r(\lambda)],
\ldots, R_{n_r}[z_1(\lambda),\ldots, z_r(\lambda)]$
be a system of Riccatians associated with a simple Lie algebra ${\cal L}_r$
and represented in a canonical form.  Then the solution of
problem 2.1 has the following form:
\begin{eqnarray}
z_a(\lambda) &=& F_a(\lambda)+
\sum_{i=1}^{M_a}\frac{1}{\lambda - \xi_{i}^a},
\label{5.11}\ea\label{5.11a}\\
c_{n_a}(\lambda) &=& R_{n_a}\left[F_1(\lambda)+
\sum_{i=1}^{M_1}\frac{1}{\lambda - \xi_{i}^1},\ldots, F_r(\lambda)+
\sum_{i=1}^{M_r}\frac{1}{\lambda - \xi_{i}^r} \right],
\eb\label{5.11b}
\end{eqnarray}
where $M_a, \ a=1,\ldots, r$ are arbitrary non-negative
integers, and the parameters $\xi_{i}^a, \ i=1,\ldots, M_a,
\ a=1,\ldots, r$ satisfy the system of equations
\begin{eqnarray}
F^a(\xi_{i}^a)+
\sum_{b}\sum_{k=1}^{M^b}\frac{(\pi^a,\pi^b)}
{\xi_{i}^a - \xi_{k}^b}=0, \quad i=1,\ldots, M_a,
\ a=1,\ldots, r.
\label{5.12}
\end{eqnarray}
In particular, the eigenvalues $c_2(\lambda)$ are given by
the formula
\begin{eqnarray}
c_2(\lambda) &=& \sum_{a,b=1}^r (\pi^a,\pi^b)
\left(F_a(\lambda)+\sum_{i=1}^{M_a}\frac{1}{\lambda-\xi^a_i}\right)
\left(F_b(\lambda)+\sum_{i=1}^{M_b}\frac{1}{\lambda-\xi^a_i}\right)+
\nonumber\\
&+& \sum_{a=1}^r (\pi^a,\pi^a)
\left(F_a(\lambda)+\sum_{i=1}^{M_a}\frac{\hbar}{\lambda-\xi^a_i}\right)'.
\label{5.212}
\end{eqnarray}
For any set of numbers $M_1,\ldots, M_r$ the
equations (\ref{5.12}) have a finite set of solutions and
therefore, the spectrum of the generalized Gaudin problem
is infinite and discrete}.

{\bf Proof.} The proof of this theorem immediately follows
from general theorem 2.1 and concrete forms of regularizing
polynomials and second-order Riccatians given by formulas (\ref{5.9})
and (\ref{5.10}).

\medskip
In next section we demonstrate that this solution exactly
coincides with solution of the generalized Gaudin spectral
problem associated with algebra ${\cal L}_r$.

\section{Generalized Gaudin spectral problem}

In this section we consider Gaudin models associated with
arbitrary simple Lie algebras. Let ${\cal L}_r$ be a simple
Lie algebra of rank $r$ and dimension $d$. The corresponding
Gaudin algebra, which we denote by  ${\cal G}[{\cal L}_r]$, is an
infinite-dimensional extension of algebra ${\cal L}_r$. Its
covariant generators we denote by $S^A(\lambda)$, where $\lambda$
is a complex parameter playing the role of a continuous
index, and $A$ is a discrete index.  The commutation
relations for these operators can be written in the form
\begin{eqnarray}
[S^A(\lambda),S^B(\mu)]=-\hbar {C^{AB}}_C
\frac{S^C(\lambda)-S^C(\mu)}{\lambda-\mu},
\label{4.12}
\end{eqnarray}
in which ${C^{AB}}_C$ are structure constants of algebra
${\cal L}_r$.

The Cartan--Weyl decomposition of algebra ${\cal L}_r$
induces an analogous decomposition of the Gaudin algebra
${\cal G}[{\cal}_r]$. Correspondingly,
the set of $d$ generators $S^A(\lambda)$
can be divided into three subsets consisting of raising and lowering
operators, $S^\alpha(\lambda), \ \alpha\in R_{\pm r}$,
associated with positive and negative roots of algebra
${\cal L}_r$ and $r$ neutral operators $S^a(\lambda), \ a\in N_r$.

The lowest weight representations of algebra ${\cal G}[{\cal
L}_r]$ are determined by the formulas
\begin{eqnarray}
S^a(\lambda)|0\rangle=F^a(\lambda)|0\rangle, \quad a\in N_r,
\nonumber\\
S^\alpha(\lambda)|0\rangle=0, \quad \alpha\in R_{-r},
\label{4.103}
\end{eqnarray}
where $|0\rangle$ is the lowest weight vector and
$F^a(\lambda)$ are covariant components of the lowest
weight $\vec F(\lambda)$.
The representation space is then defined as
\begin{eqnarray}
W_{\vec F(\lambda)}= \mbox{linear span of vectors} \
\{S^{\alpha_1}(\lambda_1)\c
dots
S^{\alpha_n}(\lambda_n)|0\rangle\}
\label{4.103a}
\end{eqnarray}
with arbitrary
$\alpha_1,\ldots,\alpha_n \in R_{+r}$ and $\lambda_1,\ldots,\lambda_n$
for each $n=0,1,2,\ldots$.

Consider the operator
\begin{eqnarray}
C_2(\lambda)= g_{AB}S^A(\lambda)S^B(\lambda)
\label{4.14}
\end{eqnarray}
which belongs to the universal enveloping algebra of
algebra ${\cal G}[{\cal L}_r]$ and has the form similar to
the form of the second-order Casimir operator for algebra
${\cal L}_r$. As before, we call it the {\it
Casimir--Gaudin operator}. Not being a Casimir invariant for
algebra ${\cal G}[{\cal L}_r]$, the operator $C_2(\lambda)$
forms however a commutative family
\begin{eqnarray}
[C_2(\lambda),C_2(\mu)]=0.
\label{4.15}
\end{eqnarray}
This property suggests to interpret $C_2(\lambda)$ as a
generating function of commuting integrals of motion for
some quantum system. It can be shown that,
in contrast with the $sl(2)$ case, the functions $C_2(\lambda)$
do not contain enough number of commuting integrals of
motion sufficient for claiming that the corresponding
quantum system is completely integrable. The question of a
complete integrability of these quantum problems is still open
except the problems associated with algebra $sl(3)$ (see
discussion in section 7).  Fortunately, irrespective of a concrete
answer to this question, one can demonstrate that the spectral
problem for them admits a simple and elegant solution.
Taking the representation space
of Gaudin algebra ${\cal G}[{\cal L}_r]$ as an analog of
the space of states, we can formulate the following analog
of the Schr\'odinger problem for operators $C_2(\lambda)$:

\medskip
{\bf Problem 6.1.} {\it Find all solutions of the spectral equation
\begin{eqnarray}
C_2(\lambda)\phi=c_2(\lambda)\phi, \quad \phi \in W_{\vec F(\lambda)},
\label{4.21}
\end{eqnarray}
provided that the lowest weight
$F(\lambda)=F^1(\lambda),\ldots, F^r(\lambda)$ is given.}

\medskip
{\bf Definition 6.1.} The equation (\ref{4.21}) is called
the {\it ${\cal L}_r$ Gaudin spectral equation} and
the models described by ``hamiltonians'' $C_2(\lambda)$ we
refer to as {\it ${\cal L}_r$ Gaudin models}.

\medskip
{\bf Theorem 6.1.} {\it The solution of problem 6.1 for algebras
$A_r, B_r, C_r, D_r, E_6$ and $E_7$ has the following form
\begin{eqnarray}
\phi&=&\prod_{a=1}^r\prod_{i=1}^{M_a}S^{\pi^a}(\xi^a_{i})|0\rangle
+ \ldots
\label{4.211}\ea\label{4.211a}\\
c_2(\lambda) &=& \sum_{a,b=1}^r (\pi^a,\pi^b)
\left(F_a(\lambda)+\sum_{i=1}^{M_a}\frac{1}{\lambda-\xi^a_i}\right)
\left(F_b(\lambda)+\sum_{i=1}^{M_b}\frac{1}{\lambda-\xi^a_i}\right)+
\nonumber\\
&+& \sum_{a=1}^r (\pi^a,\pi^a)
\left(F_a(\lambda)+\sum_{i=1}^{M_a}\frac{1}{\lambda-\xi^a_i}\right)'.
\eb\label{4.211b}
\end{eqnarray}
where $\xi_i^a$ are the numbers satisfying the system of equations
\begin{eqnarray}
\sum_{b=1}^r \sum_{k=1, k\neq i}^{M_b}\frac{(\pi^a,\pi^b)}
{\xi^a_i-\xi^b_k}=F^a(\xi^a_i),
\label{4.213}
\end{eqnarray}
where $a=1,\ldots, r$ and $i=1,\ldots, M_a$.
For any set of numbers $M_1,\ldots, M_r$ the
equations (\ref{4.213}) have a finite set of solutions and
therefore, the spectrum of the generalized Gaudin problem
is infinite and discrete}.

\medskip
The proof of this theorem can be found in refs. [Ushveridze
1990, 1992, 1994].
For the proof for only classical Lie algebras $A_r,B_r,C_r,D_r$
see also [Jur\v{c}o 1989]. As far as we know, the solutions of the Gaudin
spectr
al
problems associated with three exceptional Lie algebras $G_2,
F_4, E_8$ are not yet found.

\section{Conclusion}

Comparing the results of sections 5 and 6 we can make sure
that the spectrum of the Riccatian $R_{2}[{\bf z}(\lambda)]$
associated with algebra ${\cal L}_r$ exactly coincides with
the spectrum of the second-order Casimir -- Gaudin operator
$C_2(\lambda)$ constructed from generators of Gaudin
algebra ${\cal G}[{\cal L}_r]$. Moreover, the Riccatian
$R_{2}[{\bf z}(\lambda)]$ itself has the same form as the
Cartan part of the normally-ordered\footnote{Normal
ordering means that all lowering operators are transfered
to the right.} operator $C_2(\lambda)$. We can go further
and show that the higher Riccatians $R_{n_i}[{\bf z}(\lambda)]$
also coincide with Cartan parts of normally-ordered
higher-order Casimir -- Gaudin operators  $C_{n_i}(\lambda)$ (i.e. operators
constructed from generators of Gaudin algebra
${\cal G}[{\cal L}_r]$  and having the same structure
as independent Casimir invariants of algebra ${\cal L}_r$.)

Evidently, all these coincidences cannot be accidental and
should manifest of a deep relationship between Riccati and
Gaudin spectral problems.

Two thinks immediately come in ones head. First is that the generalized
Gaudin models are integrable and the operators
$C_{n_i}(\lambda), \ i=1,\ldots, r$ (or their slight
deformations) form a complete set
of quantum integrals of motion\footnote{As in the classical
case.} whose spectra coincide with spectra of Riccatians.
Second think is that the Gaudin models admit separation of variables and the
resulting multi-parameter spectral equations appearing
after the separation are somehow related to generalized
Riccati equations\footnote{As in the $sl(2)$ case}.

It turns out, however, that the verification of these assumptions
is far from being a simple matter.
The main difficulty lies in the fact that the operators $C_{n_1}(\lambda),
\ldots, C_{n_r}(\lambda)$ do not generally form commutative
families\footnote{Except the cases of algebras $A_1$ and $A_2$},
and thus, cannot be considered as generating functions for
quantum integrals of motion\footnote{For example, this can be easily
demonstrated by direct calculation of commutators for
algebra $A_3$}. There are two possible explanations of this situation:
1) the Gaudin models are non-integrable on the quantum level,
and 2) they are integrable but the form of additional integrals of
motion is more complicated\footnote{The method for
constructing additional integrals of motion for quantum models
associated with sl(n) solutions of Yang -- Baxter equation,
given in [Kulish and Sklyanin 1982], is wrong because it is based on the use of
non-invertible R-matrices.}.

In our opinion, the second possibility seems more
realistic, and in order to proof this, we can start with the systems of
Riccati equations and try to obtain the Gaudin models by
means of the inverse method of separation of variables
(exactly in the same way as for the $A_1$ case).
This program will be realized in next publication. Here we
only anonce some preliminary results which, in our opinion,
are rather interesting.

\medskip
$\bullet$ The systems of generalized Riccati equations
can be separated, i.e. transformed into systems of
$r$ different equations, each of which contains only one of functions
$z_1(\lambda),\ldots, z_r(\lambda)$.

$\bullet$ The separated systems can be linearized
by means of the substitution $z_i(\lambda)=\psi_i'(\lambda)/\psi_i(\lambda)$,
after which one obtains a system of linear multi-parameter
spectral equations.

$\bullet$ The main feature of obtained equations is
that they depend on spectral parameters {\it non-linearly}
(except the cases of algebras $A_1$ and $A_2$).

$\bullet$ As a consequence of this non-linearity,
the completely integrable models obtained from these
multi-parameter spectral equations
after applying to them the inverse method of separation of variables
are non-linear in the sense
that, along with ordinary (linear) integrals of motion, they
contain a set of nonlinear ones, represented by non-linear operators
in Hilbert space! Evidently, such models cannot be
considered as ordinary quantum models.

$\bullet$ The integrals of motion for these models, which
we denote by $\hat C_{n_i}(\lambda)$, have the same spectra as the
Riccatians $R_{n_i}[{\bf z}(\lambda)]$.

\medskip
Unfortunately, up to now it is not clear whether these
``pseudo-quantal'' completely integrable models do actually have some
relation to Gaudin models. In order to prove this
we should try to demonstrate that the linear operator
$\hat C_{2}(\lambda)$ constructed in such a way is
nothing else than the second-order Casimir --
Gaudin operator $C_{2}(\lambda)$. If it will be done then one can claim
that the remaining integrals
of motion of Gaudin models are some {\it non-linear}
deformations of high-order Casimir -- Gaudin operators!
We hope to return to this interesting question in next publication.

\medskip
In conclusion I would like to thank my colleagues from the Theoretical
Department for many valuable comments and remarks.

\newpage

\section{References}

\medskip
Gaudin M 1976 {\it J. Physique} {\bf 37} 1087--98

\medskip
\noindent  Gaudin M 1983 {\it La Fonction d'Onde de Bethe} (Paris: Masson)

\medskip
\noindent Jur\v{c}o V E 1989 {\it J. Math. Phys.} {\bf 30} 1289--91

\medskip
\noindent Korn G A and Korn T M 1971 {\it Mathematical
Handbook} (New York: McGraw-Hill)

\medskip
\noindent Kulish P P and Sklyanin E K 1982
{\it Lect. Not. Phys.} {\bf 151} 61--119

\medskip
\noindent Sklyanin E K 1987 {\it Zap. Nauchn. Semin. LOMI} {\bf 164} 151--170
(in Russian)

\medskip
\noindent Sklyanin E K 1991 Preprint of Helsinki University
HU-TFT-91-51, Helsinki (see also hep-th/9211111)

\medskip
\noindent Sklyanin E K 1992 Preprint of Cambridge University
NI-92013, Cambridge

\medskip
\noindent Ushveridze A G 1989 {\it Sov. J. Part. Nucl.} {\bf 20} 1185--245

\medskip
\noindent Ushveridze A G 1990 Preprint of Georgian Institute of
Physics FTT-16, Tbilisi (in Russian)

\medskip
\noindent Ushveridze A G 1992 {\it Sov. J. Part. Nucl.} {\bf 23} 25--51

\medskip
\noindent Ushveridze A G 1994 {\it Quasi-exactly solvable models in
quantum mechanics} (Bristol: IOP Publishing)

\end{document}